\documentclass[a4paper]{article}
\usepackage{booktabs} 

\usepackage[ruled]{algorithm2e} 
\usepackage{float}
\usepackage{listings}
\usepackage{tikz}
\usepackage{standalone}
\usetikzlibrary{shapes,arrows,positioning,snakes,calc,arrows.meta,backgrounds}
\tikzstyle{block} = [draw, rectangle, 
minimum height=3em, minimum width=6em]
\tikzstyle{smblock} = [draw,rectangle, minimum height=3em,minimum width=3em]
\tikzstyle{longblock} = [draw,rectangle, minimum height=3em,minimum width=9em]
\tikzstyle{longblock2} = [draw,rectangle, minimum height=1.4em,minimum width=9em]
\tikzstyle{longblock2sm} = [draw,rectangle, minimum height=1.4em,minimum width=3em]
\tikzstyle{ssmblock} = [draw,rectangle, minimum height=2em,minimum width=2em]
\tikzstyle{tallblock} = [draw, rectangle, minimum height=10em, minimum width=4em]
\tikzstyle{talltallblock} = [draw, rectangle, minimum height=16em, minimum width=4em]
\tikzstyle{carblock} = [draw, rectangle, minimum height=6em, minimum width=3em]
\tikzstyle{smwideblock} = [draw, rectangle, minimum height=2em, minimum width=3em]
\tikzstyle{bus} = [draw, rectangle, minimum height=2em, minimum width=0.5em]
\tikzstyle{sum} = [draw,  circle, node distance=1cm]
\tikzstyle{newsum} = [draw, circle]
\tikzstyle{disk} = [draw, circle, fill=black, minimum size=0.1cm, inner sep=0pt]
\tikzstyle{input} = [coordinate]
\tikzstyle{output} = [coordinate]
\tikzstyle{pinstyle} = [pin edge={ <-,black}]
\tikzstyle{square2} = [draw,rectangle, minimum height=0.5em,minimum width=0.5em,fill=black]
\tikzstyle{square3} = [draw,rectangle, minimum height=0.25em,minimum width=0.25em,fill=black]
\tikzstyle{wheelblock} = [rectangle,minimum height=1em, minimum width=0.1em]
\tikzstyle{circle1} = [draw, circle, minimum size=3em, inner sep=4pt]
\tikzstyle{circle2} = [draw, circle, minimum size=0.75em, inner sep=0pt, fill=black]
\tikzstyle{roscircle} = [draw, circle, minimum size=1em, inner sep=0pt]
\usepackage{tikzscale}
\usepackage{adjustbox}
\usepackage{pgfplots}
\pgfplotsset{compat=newest}
\usetikzlibrary{plotmarks,fit,spy}
\usetikzlibrary{arrows.meta}
\usepgfplotslibrary{patchplots}
\usepackage{grffile}
\usepackage{amsmath}
\usepackage{textcomp}
\usepackage[skins]{tcolorbox}

\usepackage[caption=false]{subfig}

\def\addlegendimage{\csname pgfplots@addlegendimage\endcsname}

\pgfplotsset{
	legend image with text/.style={
		legend image code/.code={%
			\node[anchor=center] at (0.3cm,0cm) {#1};
		}
	},
}

\pgfplotsset{plot coordinates/math parser=false}
\newlength\figureheight
\newlength\figurewidth

\usepackage{pdfpages}

\newcounter{image}
\tikzset{
	square/.style={
		rectangle,
		draw=black,
		minimum width=1cm,
		minimum height=1cm,
		text centered
	},
	drawinside/.code args={#1}{
		\draw($(#1.west)!0.3!(#1.center)$)--($(#1.east)!0.3!(#1.center)$);
		\draw($(#1.south)!0.3!(#1.center)$)--($(#1.north)!0.3!(#1.center)$);
		\draw($(#1.south west)!0.4!(#1.west)!0.3!(#1.center)$)--($(#1.south west)!0.165!(#1.west)!0.5!(#1.center)$)--(#1.center);
		\draw(#1.center)--($(#1.north east)!0.165!(#1.east)!0.65!(#1.center)$)--($(#1.north east)!0.45!(#1.east)!0.45!(#1.center)$);            
	},
	record/.style args={#1 and #2}{
		rectangle,draw,minimum width=#1, minimum height=#2
	}
}

\pgfplotsset{every axis/.append style={
		tick label style={font=\LARGE}  
	}, every axis label/.append style={font=\LARGE}}

\PassOptionsToPackage{hyphens}{url}
\usepackage{hyperref}

\SetAlFnt{\small}
\SetAlCapFnt{\small}
\SetAlCapNameFnt{\small}
\SetAlCapHSkip{0pt}

\IncMargin{-\parindent}

\begin{document}
\title{Low Cost, Open-Source Testbed to Enable Full-Sized Automated Vehicle Research}  

\author{Austin~Costley \and Chase Kunz \and Ryan Gerdes \and Rajniaknt Sharma }

\maketitle

\begin{abstract}
An open-source vehicle testbed to enable the exploration of automation technologies for road vehicles is presented. The platform hardware and software, based on the Robot Operating System (ROS), are detailed. Two methods are discussed for enabling the remote control of a vehicle (in this case, an electric 2013 Ford Focus). The first approach used digital filtering of Controller Area Network (CAN) messages. In the case of the test vehicle, this approach allowed for the control of acceleration from a tap-point on the CAN bus and the OBD-II port. The second approach, based on the emulation of the analog output(s) of a vehicle's accelerator pedal, brake pedal, and steering torque sensors, is more generally applicable and, in the test vehicle, allowed for the full control vehicle acceleration, braking, and steering. To demonstrate the utility of the testbed for vehicle automation research, system identification was performed on the test vehicle and speed and steering controllers were designed to allow the vehicle to follow a predetermined path. The resulting system was shown to be differentially flat, and a high level path following algorithm was developed using the differentially flat properties and state feedback. The path following algorithm is experimentally validated on the automation testbed developed in the paper.
\end{abstract}

%
%


%
%


\section{Introduction}
In recent years, the automotive industry has been automating vehicle systems to aid drivers with features such as adaptive cruise control, lane keeping, and collision avoidance \cite{fortune1}. In more advanced systems, that are not commercially available, vehicles can drive themselves to user defined destinations \cite{guizzo1}. Though these driver aid systems are just starting to emerge in production vehicles, research has been conducted for many years to help develop this technology \cite{rajamani}. The benefits of automated vehicles extends beyond convenience, and includes, safer roadways, increased highway throughput, and reduced emissions. Despite these positive effects of vehicle automation, there are possible drawbacks, and considerations should be made regarding the safety and security of these automated systems. Hackers are constantly reviewing platforms in search of exploitable vulnerabilities, and these automated features provide attack opportunities that were previously unavailable. For example, in \cite{valasek1} and \cite{miller1}, Miller and Valasek showed that it was possible to exploit the parking assist, and lane keeping features of modern vehicles to gain limited control of acceleration and steering. In \cite{checkoway2}, Koscher et al. demonstrate the ability to disable the braking system of modern vehicles. 

\par
 A team of undergraduate and graduate researchers at Utah State University was assembled to automate a 2013 Ford Focus Electric. To complete this challenge, the team had to find a way to control the vehicle by superseding internal control modules. The resulting testbed could then be used to further research in the fields of automated vehicle control, vehicle security, and in-motion wireless power transfer research \cite{Onar1}. The Electric Vehicle and Roadway (EVR) Research Facility and Test Track \cite{evr1} at Utah State University is a facility designed to conduct in-motion wireless power transfer research, and provided lab space for this project. 



\par 
The first goal of the project was to control the vehicle without adding hardware actuators. Two approaches were discussed and prioritized in the following order: Controller Area Network (CAN) message injection, and sensor emulation. CAN message injection would attempt to take control of the vehicle by superseding modules connected to the CAN bus. Sensor emulation would supersede the user input sensors to manipulate the signals being sent to the CAN modules. A reverse engineering approach would be taken to determine if the vehicle could be controlled through either of these methods. 
\par 
Having enabled remote control of the vehicle, the utility of the testbed in allowing for automation research was demonstrated through the design and implementation of a path following control strategy.  First, system identification steps were taken to develop an accurate model of vehicle dynamics due to user input. Low level feedback controllers were developed to allow control of meaningful vehicle states, and high level controller was developed to coordinate the low level controllers to follow a desired path.

\subsection{Contributions}
The main contributions of this work are as follows:
\begin{itemize}
	\item{Affordable, open-source, experimentally validated automation testbed using the Robot Operating System (ROS) \cite{ROS}.}
	\item{Two ways to manipulate control input to the vehicle without adding external actuation hardware: CAN message injection and sensor emulation that enable the development of other vehicle testbeds.}
	\item{Software packages, hardware descriptions, and a methodological approach that can be used to remotely control nearly any vehicle, thus allowing researchers and hobbyists to explore automation techniques and vehicle security. The software packages are available at: \url{https://github.com/rajnikant1010/EVAutomation}.}
	\item{Proof-of-concept exploitation of a vehicle vulnerability wherein a compromised ECU or other device connected to the CAN bus (including via the OBD-II port) could control vehicle acceleration.}
\end{itemize}

\subsection{Literature Review}
Control of Automated Ground Vehicles (AGV's) has been a subject of research since 1955 \cite{vis06} and has naturally progressed to include commercially available passenger vehicles. In \cite{rajamani}, Rajamani details various vehicle control systems, and modeling techniques.
\par
There have been several automated vehicle competitions held to further the research in this field. The DARPA Grand Challenge, for example, was started in 2004 to encourage researchers to develop off-road autonomous vehicle technology that could be used for military applications \cite{darpa04}. The challenge was repeated in 2005 for off-road vehicles \cite{darpa05}. Then in 2007, the challenge was altered to focus on urban driving environments \cite{darpa07}. The teams in these challenges started with an existing commercial vehicle, and developed an automated system to complete the challenge. The Grand Cooperative Driving Challenge (GCDC) is another example of the advancement of automated driving through competition \cite{gcdc1}. The purpose of the GCDC is to examine cooperative automated vehicle systems. Teams develop an automated vehicle to be used in cooperative challenges with other teams. In contrast to the work by the teams in the DARPA Grand Challenge and GCDC, this work is open-source, low-cost, and has been completed without vendor support. One purpose of this work is to allow other researchers to perform vehicle automation tasks similar those in these challenges without needing support from industry. 

\par 
 Koscher et al. and Checkoway et al. in \cite{checkoway1} and \cite{checkoway2}  showed that an attacker can gain access to Electronic Control Units (ECU's) and circumvent vehicle control and safety systems. They also demonstrated that the attacker could gain access to the CAN bus remotely, and perform similar attacks. Their results included the ability to have the car ignore a brake pedal press by the driver, a complete vehicle shut down, and complete control of the visual display. The attacks however, did not allow for arbitrary control of acceleration, braking or steering, which is required for vehicle autonomy.
\par 
Miller and Valasek built on the work from Koscher and Checkoway et. al. in \cite{valasek1} and \cite{miller1}, which detail attacks on vehicle systems in a 2010 Ford Escape, 2010 Toyota Prius, and a 2014 Jeep Cherokee. They developed an extensive platform for attacking ECU's and exploiting driver assist systems. However, the attacks had limited scope for vehicle control. For example, an attack on the parking assist module was conducted on the 2010 Ford Escape, whereby, vehicle steering could be controlled when the vehicle was traveling at under 5 mph, but this attack would not work if the vehicle was moving faster. They also demonstrated the ability to cause vehicle acceleration under very specific conditions. In contrast, this work demonstrates the ability to cause arbitrary acceleration under any condition. In addition, using the approach presented here, the acceleration can be controlled from  the OBD-II port, any bus tap point, or by gaining access to an ECU. The current work also demonstrates the security concern that a vehicle could be examined and reverse engineered in a short amount of time, as it took a small team of students just under a year to develop the automated system for a commercial vehicle.
\par 
This paper is organized as follows. Section \ref{s:reverse} details the generalized reverse engineering approach used to create the testbed, as well as how the methodology naturally allows for vulnerability discovery and mitigation experimentation. The efficacy of the testbed for automation research is demonstrated in subsequent sections.  Longitudinal and lateral low level controls are discussed in Section \ref{s:lowlevel}, while Section \ref{s:path} discusses the theory of differential flatness and the path following controller. Section \ref{s:platform} gives an overview of the testbed architecture and details the hardware and software components of the automation platform. Section \ref{s:results} shows the experimental results of the low level controllers and automation, and Section \ref{s:conclusion} summarizes the findings and concludes the paper.

\section{Enabling Remote Control}\label{s:reverse}
Modern vehicles use a Controller Area Network (CAN) bus system for module-to-module communication \cite{ISO1}. Electronic Control Units (ECU's) are the CAN modules that connect to the bus that send and receive information. A CAN module receives data from sensors, processes the data, and generates the appropriate CAN message to be broadcast on the bus using an analog-to-digital operation.
\par 
The research team for the current work used the 2013 Ford Focus Electric Wiring Guide \cite{ford1} and the Auto Repair Reference Center Research Database from EBSCOhost \cite{autorepair} to understand signal path and critical connections. The wiring guide provided diagrams for most of the wires in the vehicle and included diagrams and pin-outs for the wiring connections. The Auto Repair Reference Center was particularly useful for reverse engineering the CAN protocols. It contains the CAN messages generated and received by each module, diagrams of the four CAN buses in the vehicle, and the module layout on each bus. The CAN message information was an incomplete list of general messages sent between CAN modules. For example, the list would indicate that a message about the acceleration pedal position is sent from one module to another, but it would not indicate the structure of the message, the arbitration ID, or a conversion to useful units. 
\par 
Using these resources, the team identified sensors and modules that could be used for vehicle control. Table \ref{t:sensorConnections} summarizes these findings. In addition, the team took a hands-on approach to vehicle exploration and verified the location and connections of these components. 

\begin{table*}[t]
	\centering
\resizebox{\textwidth}{!}{\begin{tabular}{l|l|l|l}
	\multicolumn{1}{c|}{\textbf{Sensor}} & \multicolumn{1}{c|}{\textbf{CAN Module}} & \multicolumn{1}{c|}{\textbf{CAN Message}} & \multicolumn{1}{c}{\textbf{CAN Arbitration ID}} \\ \hline
	Accelerator Pedal Position Sensor    & Powertrain Control Module (PCM)          & Accelerator Pedal Position  &    0x204          \\
	Brake Pedal Position Sensor          & Automatic Braking System (ABS)           & Brake Pedal Position &    0x7D                 \\
	Steering Torque Sensor               & Power Steering Control Module (PSCM)     & Steering Torque    & Unknown                      \\
	Steering Wheel Angle Sensor          & Steering Angle Sensor Module (SASM)      & Steering Wheel Angle & 0x10                  
\end{tabular}
}
	\caption{Sensor and Module Connections for Control Signals}
	\label{t:sensorConnections}
\end{table*}

\par
\begin{figure}
	\centering
	\includegraphics[width = 0.75\textwidth]{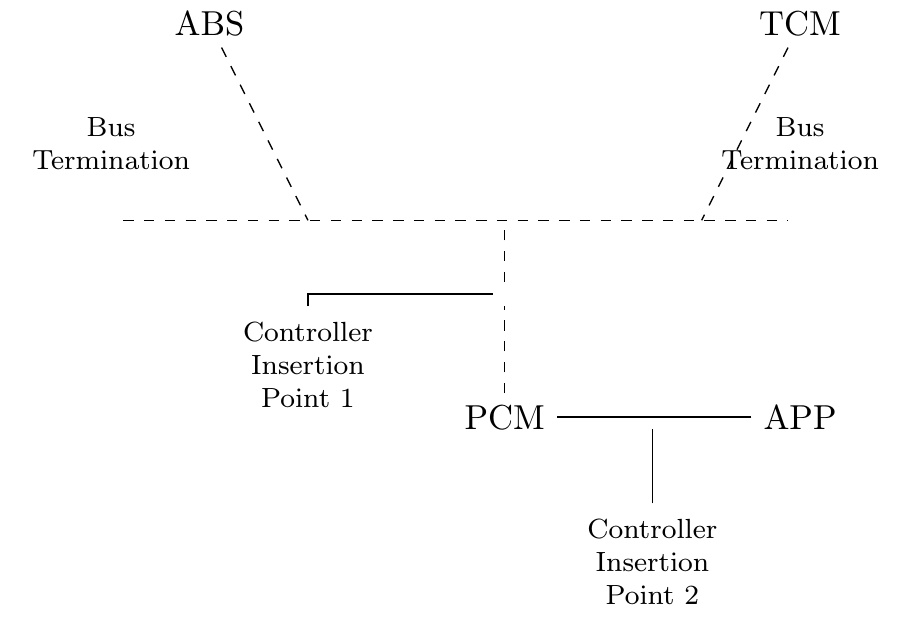}
    \caption{Vehicle CAN bus and sensor architecture. Controller insertion type 1 filters CAN messages and replaces data with the message to be injected, and is represented by a filled black square. Controller insertion type 2 emulates sensor output signals, and are represented by filled black circles. The TCM controls the main drive motor of the electric vehicle and therefore actuates acceleration. The PSCM actuates the power steering motor. The ABS module actuates the hydraulic braking system.}
	\label{f:controllerInsertion}
\end{figure}

 The examination of this architecture led to the identification of the two possible controller insertion strategies, as shown in Fig. \ref{f:controllerInsertion} . First, the CAN lines between the control module and the CAN bus could be cut, and a controller could be inserted to intercept and change messages being sent from the target control module. Second, the analog signal wires from the target sensor could be cut, and the controller could be inserted between the sensor and the control module. In either strategy, the controller would insert spurious data into the system to control the vehicle. The details and results of these two approaches are discussed in the following subsections.

\par 
In order to successfully implement the first controller insertion strategy and take advantage of the information on the vehicle CAN bus, the Vector CANalyzer \cite{canalyzer} system was used for the initial CAN message identification process. This system provides an excellent visual tool for watching CAN messages in real time. The tool displays a table of CAN messages with the rows organized by arbitration ID. The first column of the table indicates the time since the last message with a given arbitration ID was received. The second column lists the arbitration ID, and the third column shows the value for each byte of the CAN message. If a byte is changed when a new message is received, the byte is displayed in bold. Over time the byte fades to a light gray if that value stays the same. This was useful in identifying messages such as the accelerator pedal position, brake pedal position, steering wheel angle, and vehicle speed. The CANalyzer system is a great resource, but it is expensive and closed-source. Cheaper alternatives such as the Peak Systems PCAN \cite{pcan} device can be used that have an open platform for development. An open-source solution for monitoring CAN traffic with the PCAN device is provided with the open-source software accompanying this work. Further discussion on the use of the PCAN device can be found in Section \ref{ss:interface}. Another alternative is to use a microcontroller with a CAN bus interface module to monitor and report CAN traffic \cite{ti}.
\par 
Using the resources in the previous paragraphs, it was determined that CAN messages have two main functions: status and control. A status message reports the status of a vehicle component or condition, but does not control that component or condition. For example, a module will receive input from the wheel speed sensors and send the information on the CAN bus. Changing the data in this message will not result in a change of vehicle speed. A control message, however, is sent by a module to control a component or condition of the vehicle. For example, the movement of the wing mirrors is controlled by a CAN signal, when this message is changed the mirrors will move in response.

\subsection{CAN Message Injection} \label{ss:injection}

A platform for injecting CAN messages was developed using the TI TM4C129XL microcontroller evaluation kit \cite{ti} and TI CAN transceivers \cite{ti2}. The platform would connect to the CAN bus using the first controller insertion point, between the target module and the bus. The microcontroller was programmed to record and playback CAN traffic. More specifically, the microcontroller would receive the output from the control module and store it in memory. Upon user request, the output from the control module could be injected on the CAN bus. This platform was used to determine which messages, organized by arbitration ID, were generated by the target control module. Using information from the Auto Repair Center Research Database, it was determined that the Powertrain Control Module (PCM) sent a CAN message to the Transmission Control Module (TCM) regarding the accelerator pedal position. In addition, the PCM was the only control module that received the sensor signal from the accelerator pedal. For these reasons, the PCM was chosen as the target module for vehicle acceleration.

\par 

\begin{figure*}[t]
	\centering
	\subfloat[]{
		\centering
	\includegraphics[width=0.32\textwidth]{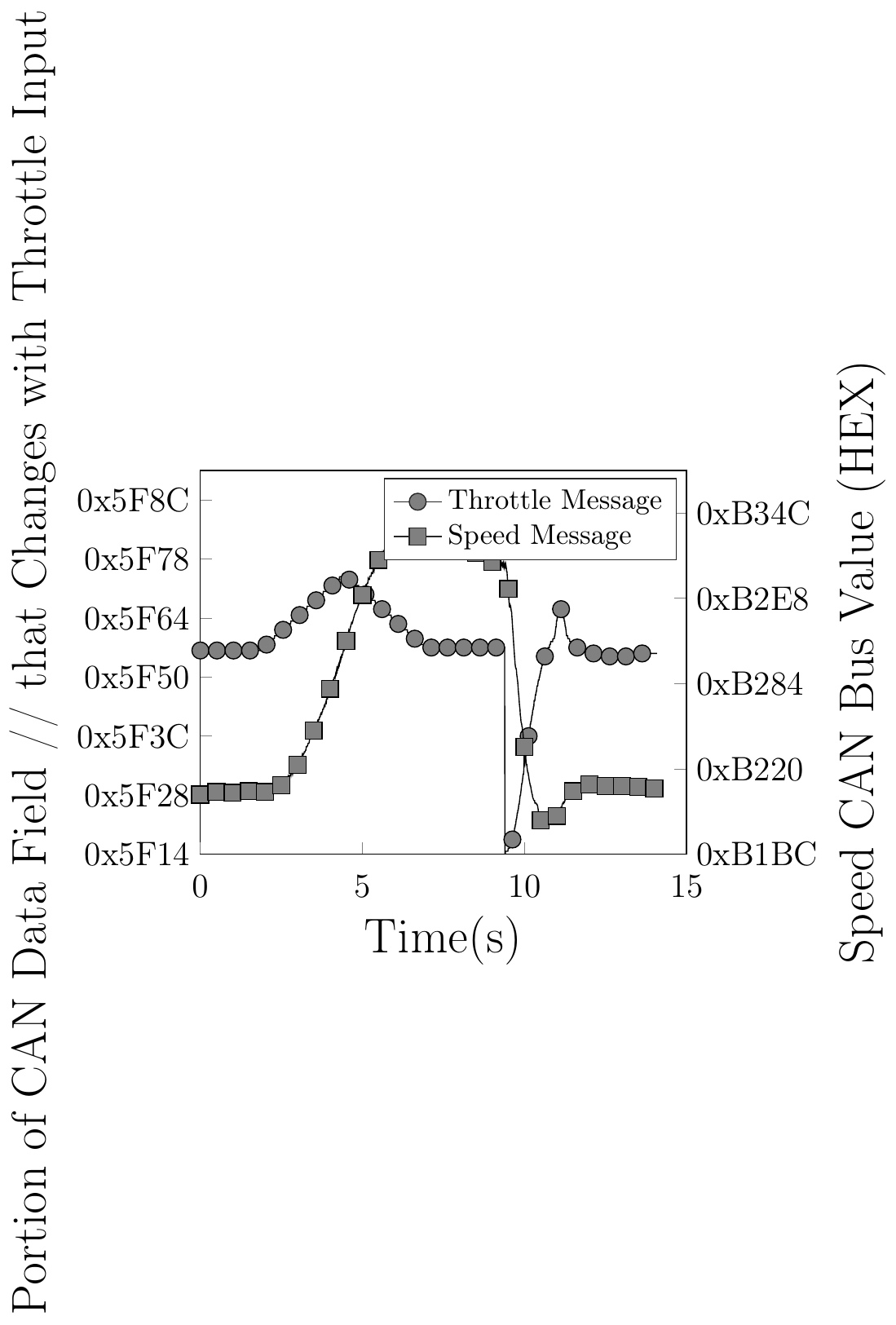}
	\label{f:CANramp}
	}
	\subfloat[]{
		\centering
	\includegraphics[width=0.32\textwidth]{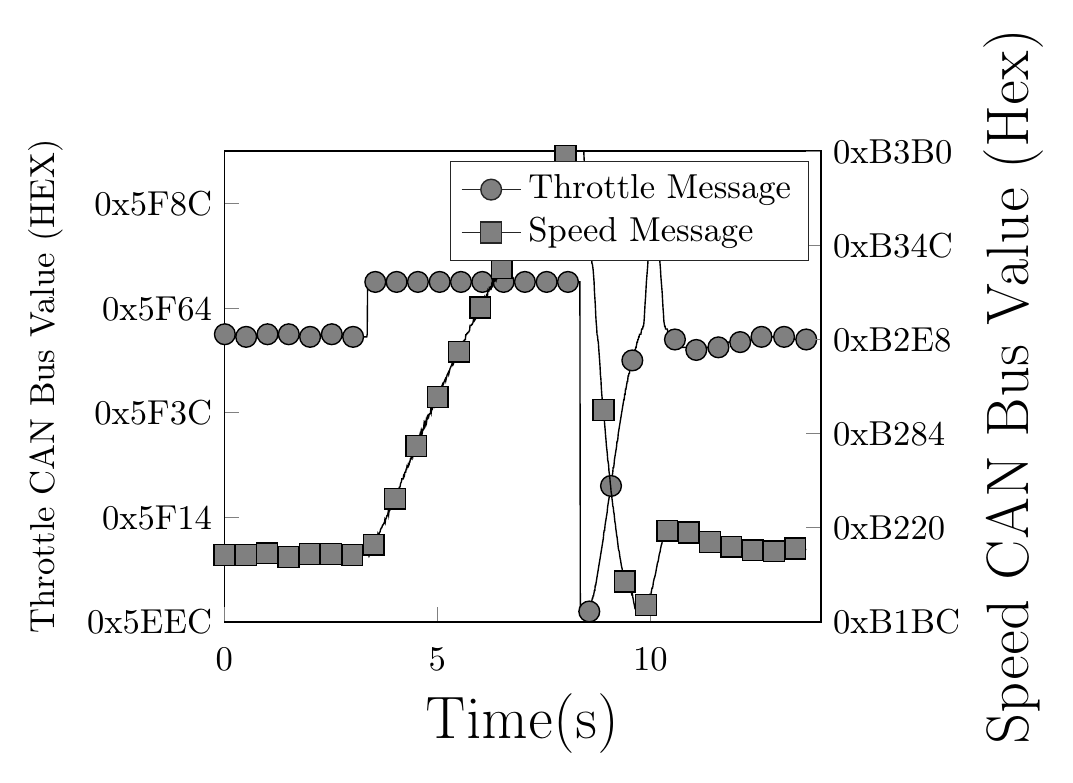}
	\label{f:CANstep}
	}
	\subfloat[]{
		\centering
	\includegraphics[width = 0.32\textwidth]{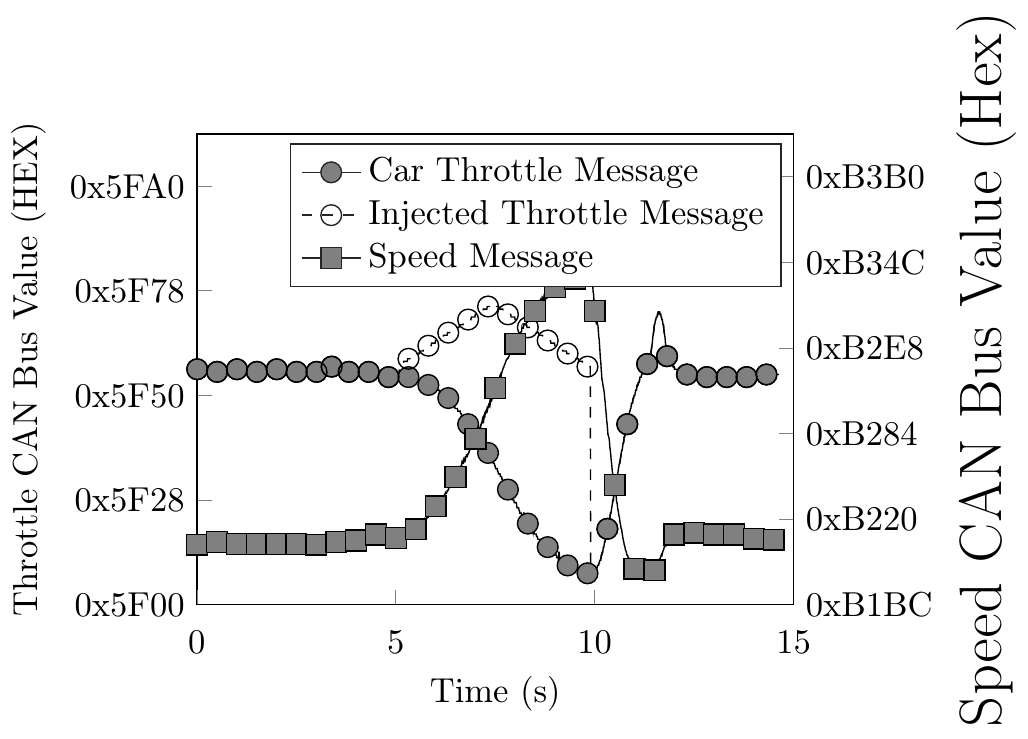}
	\label{f:obdiiramp}
	}
	
	\caption{ CAN throttle message injection from controller insertion point 1 and through OBD-II port, with resulting vehicle speed. All values read from CAN bus and represented in hexadecimal format. \protect\subref{f:CANramp} Ramp injection from controller insertion point 1.  \protect\subref{f:CANstep} Step injection from controller insertion point 1. \protect\subref{f:obdiiramp} Ramp injection through OBD-II port. The CAN bus was monitored from the OBD-II port, and the injected throttle message was broadcast on the CAN bus immediately following the reception of vehicle throttle message.} 
\end{figure*}

The connector diagrams in the Wiring Guide were used to determine the CAN bus connections to the PCM. These wires were cut and routed to the CAN injection platform. The CAN messages output from the PCM were recorded and passed through to the CAN bus for an accelerator pedal press. The recorded data was then played back on the bus and the vehicle accelerated as expected. Additional code was added to the playback function to selectively playback messages based on the message arbitration ID. This was used to search the recorded data set and isolate the acceleration control message. The messages were separated into two groups based on their arbitration ID, and each group was played back to the vehicle separately. The group that resulted in vehicle acceleration was separated again into two smaller groups and the process was repeated until one message arbitration ID was left. The acceleration control message was determined to be the message with arbitration ID \texttt{0x11A}. Figs. \ref{f:CANramp} and \ref{f:CANstep} show successful CAN injections of the acceleration control message, and the resulting speed for a ramp and step input, respectively. Due to the similarities between the acceleration control message, and a throttle signal in a gas powered car, this message is called the throttle message for the rest of this work.

\par 
Another approach to identify useful CAN messages for vehicle acceleration is by correlating the CAN messages with the vehicle speed. This approach was attempted by recording an accelerator pedal press and processing the data in Matlab. The CAN modules for the 2013 Ford Focus EV broadcast messages at prescribed frequencies as opposed to broadcasting in response to another signal. The speed message for the vehicle is broadcast at 100 Hz, which is the highest frequency messages are broadcast for this vehicle. Correlation was performed on messages of the same frequency, and the speed data was downsampled to perform correlation with messages at lower frequencies. The correlation returned a value between -1 and 1 for each byte of every message to indicate how closely correlated that byte was to the speed message. If the byte did not change during the recording the correlation returned \texttt{NAN}. On the EV-HS CAN bus there are 102 different messages and each message contains 8 bytes. The bytes were sorted based on the absolute value of the correlation value to identify the highest positively or negatively correlated bytes. The bytes that had a \texttt{NAN} value were rejected and the final number of bytes being ranked was 341. The highest correlated byte of \texttt{0x11A} was byte 4, which had a correlation value of 0.2359 and was ranked 57 out of 341. The low correlation value and rank indicates that the throttle message would not have been identified using this strategy. In contrast, using the approach described in the previous paragraph, the throttle message was identified and was used to control vehicle acceleration.

\par 
The investigation of the braking system concluded that a CAN bus message about pedal position would not actuate the hydraulic braking system. The pedal signal is sent directly to the Automatic Braking System (ABS) CAN module, which is the only CAN module on the vehicle that is connected to the hydraulic brake lines. From this it was concluded that braking could not be fully controlled  through the CAN bus.
\par 
The 2013 Ford Focus EV has an option to include park assist \cite{pam1}. Though our specific vehicle did not include this option, it was determined that the Electric Power Assisted Steering (EPAS) system had the same part number and motor as the EPAS system in a vehicle with the park assist feature. This meant that the power steering motor would be powerful enough to turn the wheel at low speeds, and by extension, any speed. The Power Steering Control Module (PSCM) receives inputs from the CAN bus and the steering torque sensor located at the base of the steering column. The torque sensor uses a torsion bar to determine the amount of torque being applied by the driver, which is used by the PSCM to determine how much assist the power steering motor should provide. Simply, for a given torque input, less assistance would be provided by the EPAS system at higher speeds. Similar to the brake system, the input of interest is sent from a sensor to the module that performs the desired action. This led to the conclusion that the control of steering would have to be controlled through torque sensor input, and not through the CAN bus. In addition, the work by Miller and Valasek \cite{miller1} \cite{valasek1} identifies the short comings of exploiting the park assist feature, namely, the park assist feature will cease to control the vehicle if the speed threshold is exceeded.

\subsection{Sensor Emulation} \label{ss:sensoremulation}
\begin{figure*}[t]
	\centering
	\subfloat[]{
		\centering
	\includegraphics[width=0.25\textwidth]{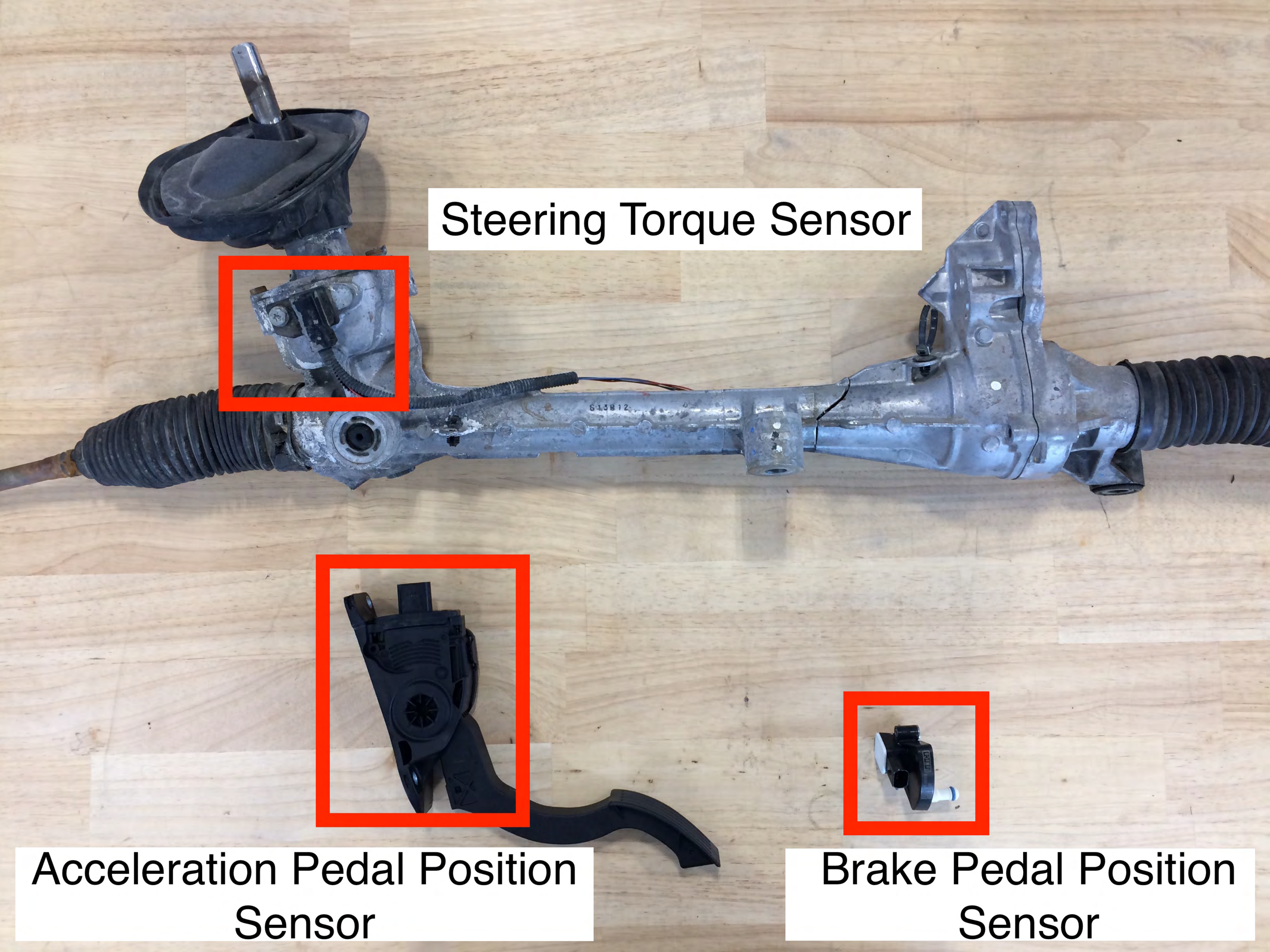}\hfill
	\label{f:sensors}
	}
		\subfloat[]{
			\centering
			\includegraphics[width=0.25\textwidth]{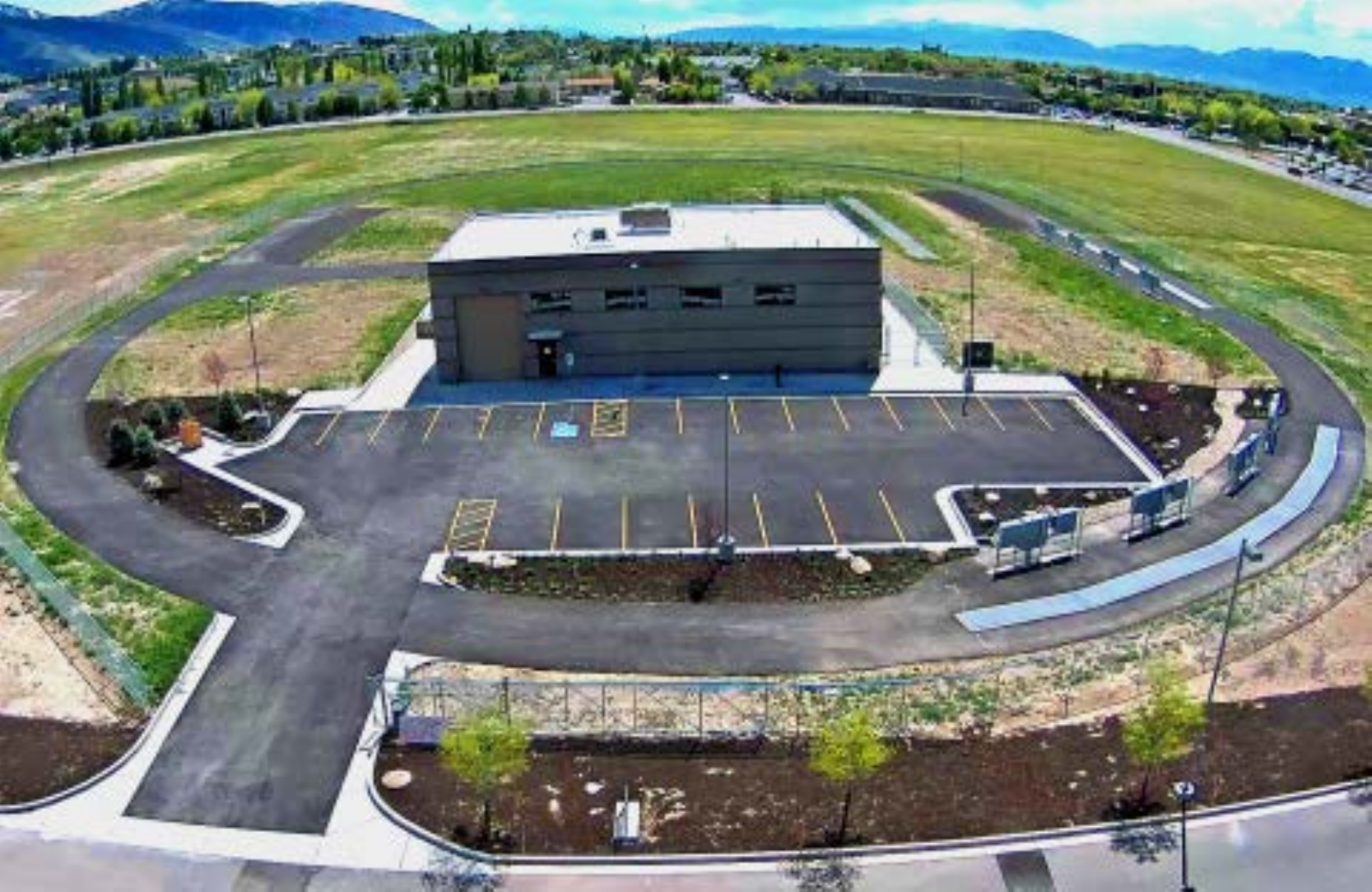}
			\label{f:evr}
		}

		\subfloat[]{
			\centering
			\includegraphics[width=0.48\textwidth]{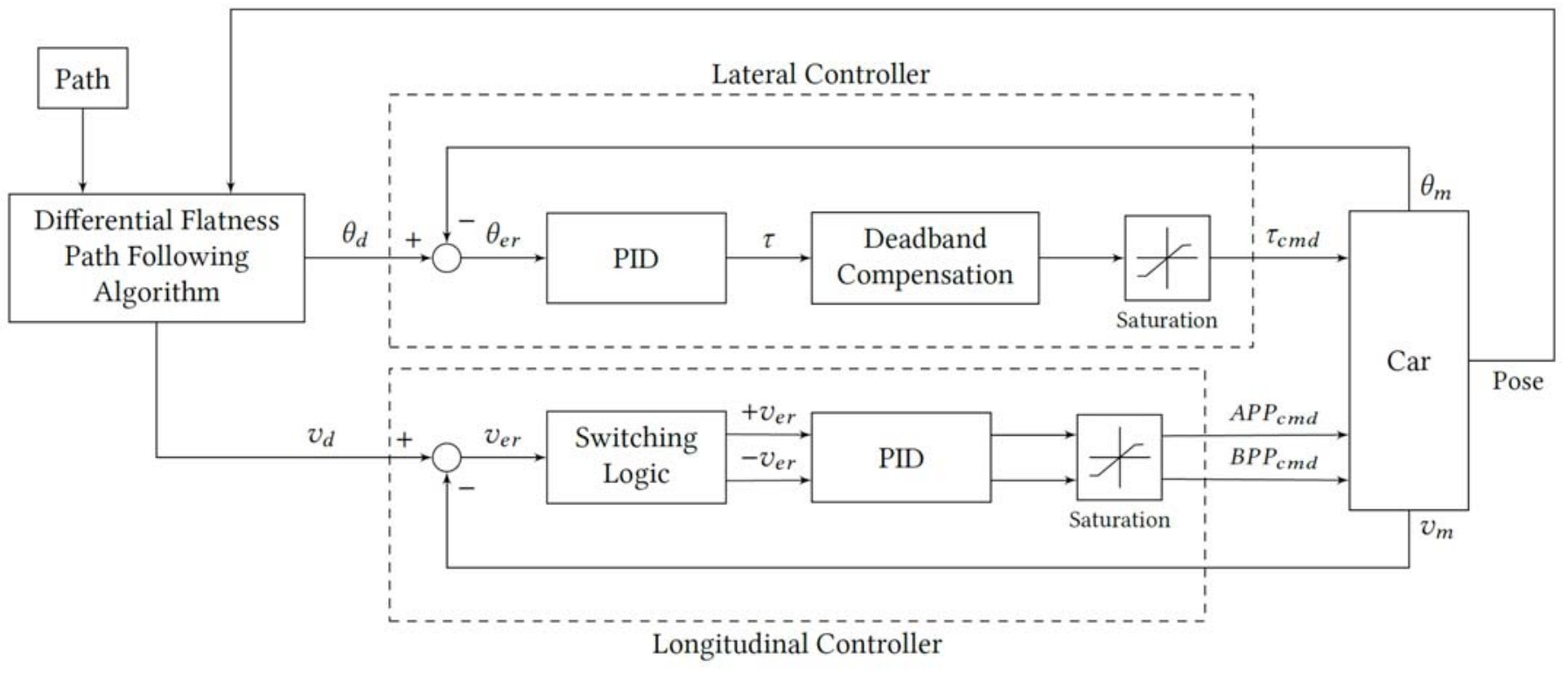}\hfill
			\label{f:high_level_diagram}
		}

	\caption{\protect\subref{f:sensors} The physical sensors emulated for vehicle control. \textit{Top:} Steering rack for 2013 Ford Focus EV, the steering torque sensor is located at the base of the steering column and measures torque from driver. \textit{Bottom Left:} The APP sensor located at the top of the accelerator pedal. \textit{Bottom Right:} The BPP sensor that is usually mounted behind the brake pedal assembly and measures the brake pedal press. \protect\subref{f:evr} Aerial view of the Electric Vehicle Roadway and Research Facility (EVR) at Utah State University.  \protect\subref{f:high_level_diagram} High-level system block diagram. Shows low-level control loop for lateral and longitudinal control and high-level differential flatness path following feedback loop.}
\end{figure*}

The CAN bus injection method was unable to control braking and steering, so the sensor emulation method was explored. The accelerator pedal position sensor was analyzed to control vehicle acceleration, the brake pedal position sensor was analyzed to control the braking system, and the steering torque sensor was analyzed to control the vehicle steering. Fig. \ref{f:sensors} shows these sensors and the following paragraphs discuss the analysis.
\par
\begin{figure*}[t]
	\centering
	\subfloat[]{
		\centering
		\includegraphics[width=0.25\textwidth]{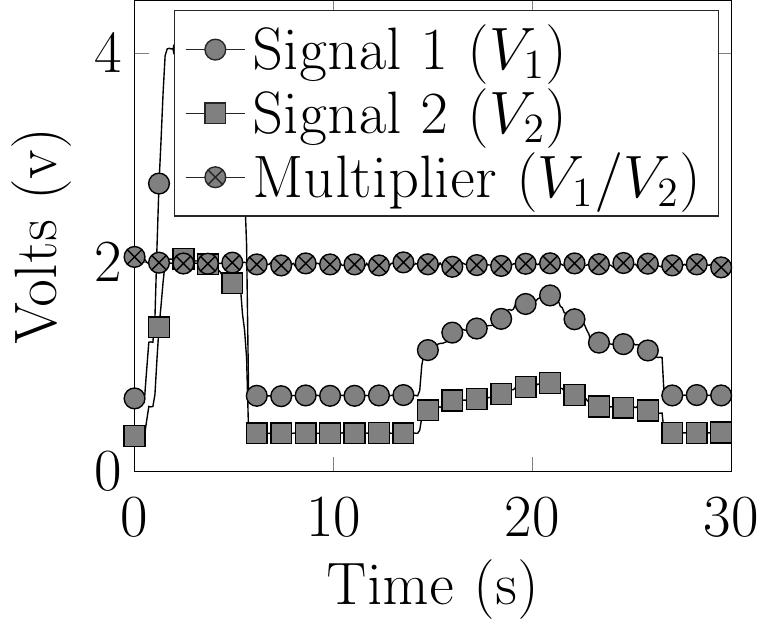}\hfill
		\label{f:app}
	}
	\subfloat[]{
		\centering
		\includegraphics[width=0.25\textwidth]{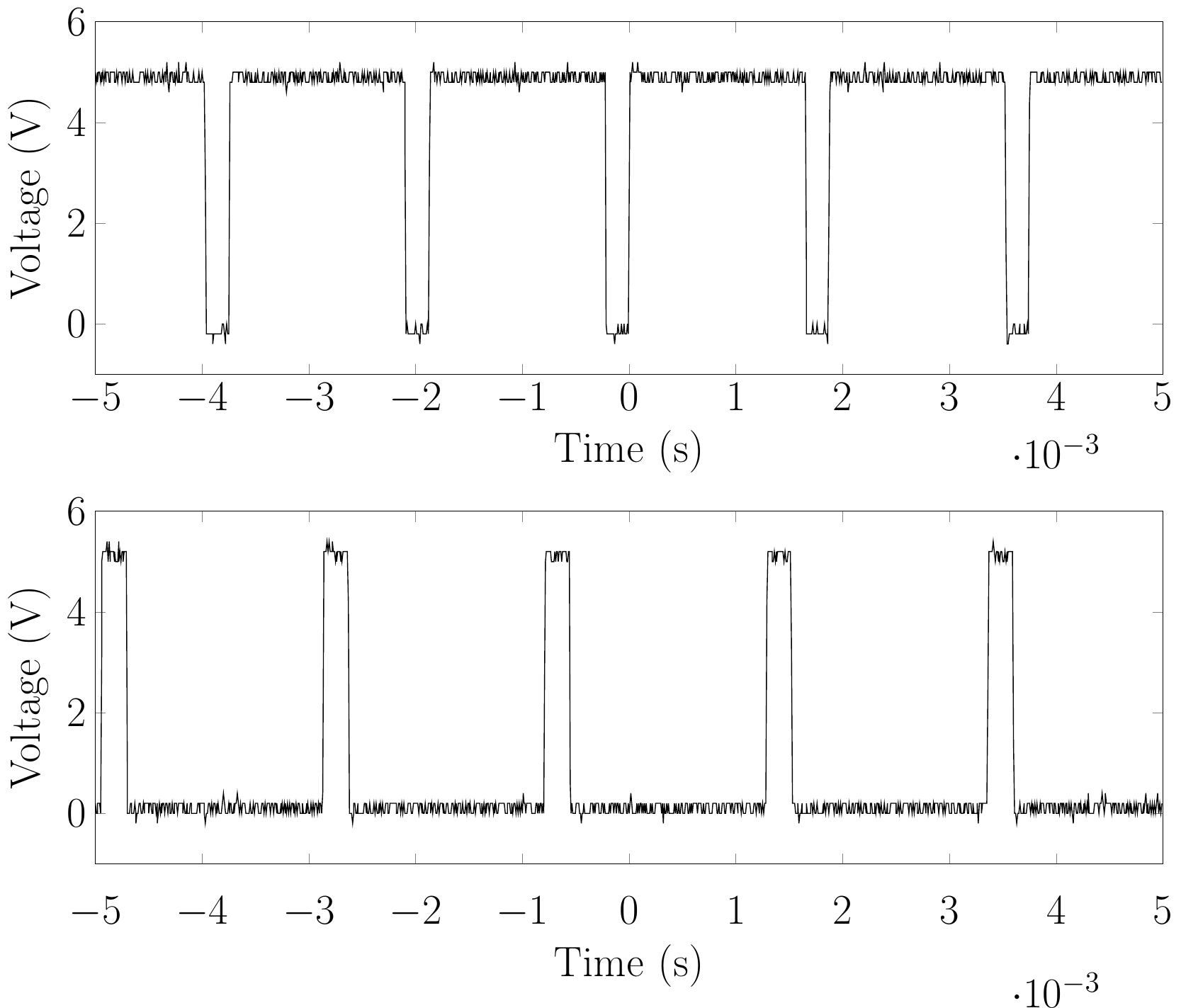}\hfill
		\label{f:bpp}
	}
	\subfloat[]{
		\centering
		\includegraphics[width=0.25\textwidth]{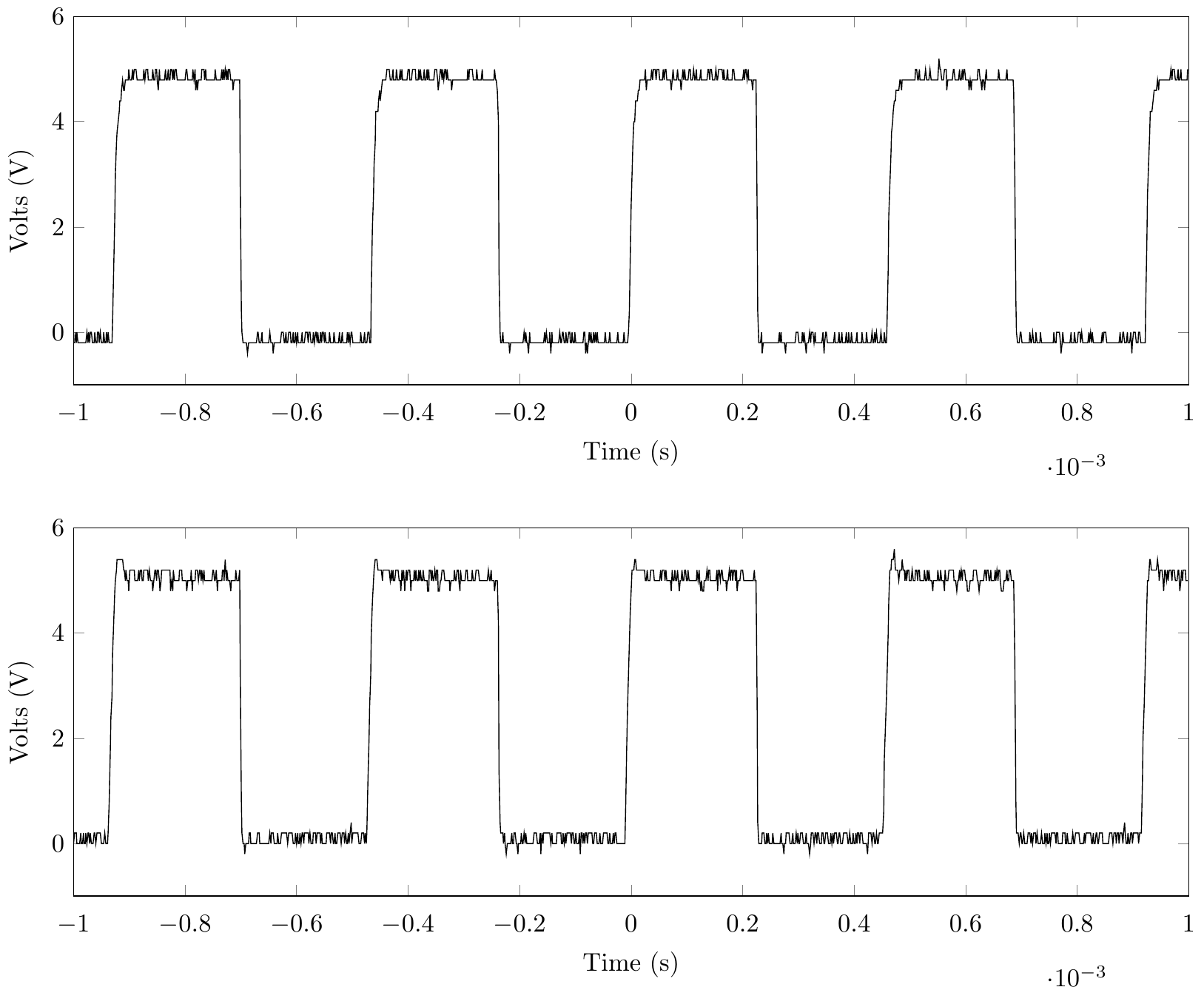}
		\label{f:steeringsigs}
	}
	
	\caption{\protect\subref{f:app} Accelerator pedal position sensor output. Two analog voltage signals related by $V_1 = 2V_2$. \protect\subref{f:bpp} Brake pedal position sensor output signals. Signal 1: 89\% resting duty cycle at 533 Hz. Signal 2: 11\% resting duty cycle at 482 Hz. \protect\subref{f:steeringsigs} Steering torque sensor output signals. 50\% resting duty cycles at 2.15 kHz.}
\end{figure*}

 
The accelerator pedal position (APP) sensor is located at the top of the accelerator pedal. There are six wires connected to the APP sensor, including two 5 V power wires with corresponding ground wires, and two signal wires. The sensor power pins were connected to a voltage source, and the signal wires were connected to an oscilloscope. It was determined that the sensor outputs two DC voltages similar to the output of potentiometers. Fig. \ref{f:app} shows the voltage levels of the two output signals in response to a pedal press. The third signal on the graph is a multiplier that relates the two signals. It is seen that $V_1 \approx 2V_2$.

\par 
The brake pedal position (BPP) sensor is located at the top of the brake pedal. There are four wires connected to the BPP, including a 5 V power, ground, and two signal wires. The BPP was connected to the voltage source and oscilloscope in the same manner as the APP. However, the BPP outputs two PWM signals instead of DC voltage levels. When the brake pedal is not pressed the duty cycles of the signals settle at 89\% for signal 1 and 11\% for signal 2. During a braking event the duty cycle for signal 1 decreases and the duty cycle for signal 2 increases at the same rate. Fig. \ref{f:bpp} shows  the two PWM signals for the BPP. The frequency of signal 1 is 533 Hz and the frequency of signal 2 is 482 Hz.

\par 
The steering torque sensor is located at the base of the steering column. The sensor connects to the Power Steering Control Module (PSCM) on the CAN bus, which determines the amount of power steering assist to provide. The assist is provided by an electric motor connected to the steering rack. In the sensor, a torsion bar is used to connect two parts of the steering shaft, where the rotational displacement can be measured to determine the torque input by the driver \cite{persson1}. Similar to the BPP, there are 4 wires connected to the steering torque sensor, including 5 V power, ground, and two signal wires. The steering torque sensor was connected to the voltage source and oscilloscope, and it was determined that the sensor outputs two PWM signals on the signal wires, where both signals settle at 50\% duty cycle when no torque is applied on the steering wheel. Both signals have a frequency of 2.15 kHz. Similar to the brake PWM signals, the duty cycles always add to 100\%, and the direction that the steering wheel is being turned determines which signal's duty cycle increases and which signal's duty cycle decreases. Fig. \ref{f:steeringsigs} shows the two steering PWM signals.

\subsection{Safety and Security}
In 1996, the OBD-II (On-Board Diagnostics) specification was required to be implemented on any new vehicle sold in the United States \cite{obdii}. This specification gives owners and technicians the ability to diagnose issues on the vehicle. The specification standardized connectors, message formats, and frequencies. The OBD-II port on the 2013 Ford Focus EV connects to the EV-HS CAN bus, which is the same bus that the throttle message is sent from the PCM to the TCM. 
\par
An attack platform was developed to inject arbitrary throttle messages through the diagnostics port. This attack method was important because, if successful, it would demonstrate that the acceleration of the vehicle could be controlled with limited intrusion. This differed from the approach in Section \ref{ss:injection}, as it does not require access to the target module, or that the CAN wires be cut and re-routed. Instead, this platform could be plugged into the OBD-II port and monitor the bus for the target message arbitration ID. Also, it would show that if an attacker was able to inject messages from any module on the EV-HS CAN bus, then arbitrary vehicle acceleration could be caused. This would stand in contrast to the findings in \cite{valasek1}, \cite{miller1}, \cite{checkoway2}, \cite{checkoway1}, where the acceleration of the vehicle could only be controlled under specific preconditions, and required intrusive access to the CAN bus.
\par 
The platform was connected to the CAN bus through the OBD-II port (other points on the bus could be used, as well) and monitored the traffic on the bus. The user determined a target message, in this case, the throttle message, and provided that message arbitration ID to the system. In Section \ref{ss:injection}, it was determined that the throttle message is included in the data frame associated with arbitration ID \texttt{0x11A}, and is broadcast at 10 Hz. The platform waited until a message was received with the corresponding arbitration ID, and would replace throttle message data with an arbitrary throttle command value. The platform was designed to only alter the parts of the message that relate to the throttle control. The inserted message would be sent 250 $\mu$s after the actual message, leaving 9.75 ms for the inserted message to be received and processed by the TCM. This allowed the inserted message to dominate the period and cause the vehicle to accelerate. Fig. \ref{f:obdiiramp} shows the successful ramp injection through the OBD-II port and the resulting vehicle speed. Thus confirming the hypothesis that vehicle acceleration can be caused by injecting CAN messages through the OBD-II port, and therefore, could be caused at any other point on the bus.

\par

These results demonstrate a CAN bus security concern. If an attacker were able to access the CAN bus, physically, or by compromising another ECU, they would be able to effect the acceleration of the vehicle without causing any errors. Remote access to the vehicle, but not necessarily the requisite CAN bus, could be effected by compromising the Telematic Control Unit (TCU) or a wireless Tire Pressure Monitoring Sensor (TPMS). The TPMS sends a signal to the Body Control Module (BCM), which in turn transmits a message on the medium speed CAN bus (MS-CAN), while the TCU is connected to the I-CAN bus (it is unlikely, however, that compromising a sensor would allow for injection of arbitrary CAN messages onto the I-CAN or MS-CAN bus). These busses are connected to the EV-HS bus through a gateway module; transmitting a message from one bus to another, which would be required for either the TCU or TPMS to impersonate the PCM by passing APP messages, was not explored in this work. Regardless of the access approach, the driver is able to stop the unwanted acceleration by pressing the brake pedal, however, other works indicate that it is possible to make the vehicle ignore braking requests \cite{miller1} \cite{valasek1} \cite{checkoway1} \cite{checkoway2}. This was not investigated as part of this work. Another security concern is that of a malicious technician. Since technicians will often access the OBD-II port when a vehicle is being serviced, it would be quite simple for them to leave an OBD-II injection platform connected to the OBD-II port. The acceleration control could be initiated remotely or by a timer, causing the vehicle to accelerate at a dangerous time.

\par 
We present two remediation strategies that could be employed to help protect against this vulnerability. First, a simple change in the acceleration system architecture, such that the APP sensor connects  directly to the TCM, which is the actuating module. This would remove the need of a throttle message to be sent from the PCM to the TCM and effectively remove the attack surface. The second approach is through device fingerprinting for both the digital and analog signals \cite{murvay1} \cite{cho1}. This would allow the receiving module to authenticate the transmitting module, and prevent this type of attack.

\section{Low Level Control} \label{s:lowlevel}
A simple overview of the control structure for the automated vehicle platform is shown in Fig. \ref{f:high_level_diagram}. The high level controller plans the path and provides the desired vehicle speed, $v_{desired}$, and desired steering wheel angle, $\theta_{desired}$, discussion on the high level controller can be found in Section \ref{s:path}. The low level controllers discussed in this section are the inner loops that control vehicle speed and steering wheel angle. The vehicle commands are $\tau_{cmd}$, $APP_{cmd}$, and $BPP_{cmd}$, and represent, steering torque, APP, and brake pedal position, respectively. 
\par 
The first step in the development of the low level controllers was to determine a model of the system being controlled. A system model is expressed as a transfer function relating the input to the output of the system. Models were identified to relate the accelerator and brake pedal inputs to vehicle velocity, and steering wheel angle to vehicle heading. The following subsections review the model identification approach, and low level controller design process for the Ford Focus.

\subsection{Longitudinal Model}
The longitudinal characteristics of the vehicle are affected by the APP sensor, and the BPP sensor. These two systems were tested and identified separately, then implemented together as a complete longitudinal model. 
\par
In \cite{dias15}, Dias et al. perform longitudinal model identification and controller design for an autonomous vehicle. This approach was examined for the current work, however, a more straightforward classical controls technique using step responses was ultimately used. Once the acceleration and braking systems were identified, a control system was developed for each input device. The control loops for accelerator and braking were connected by switching logic to determine whether a the accelerator or brakes should be used. A similar two loop control system with a switching logic component was used for the longitudinal controller in the current work. However, this work is an open-source project that uses the Robot Operating System (ROS) \cite{ROS}.
\par
\begin{figure*}[t]
	\centering
	\subfloat[]{
		\centering
		\includegraphics[width=0.24\textwidth]{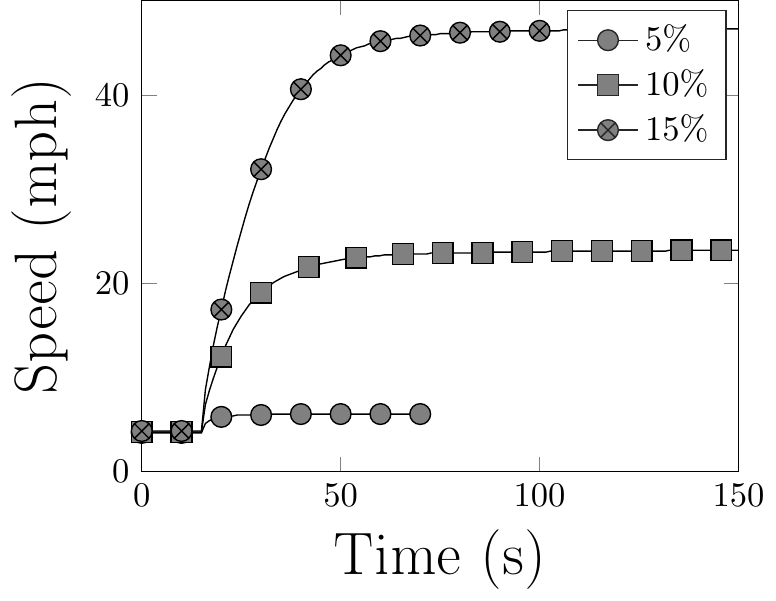}\hfill
		\label{f:appstep}
	}
	\subfloat[]{
		\centering
	\includegraphics[width=0.24\textwidth]{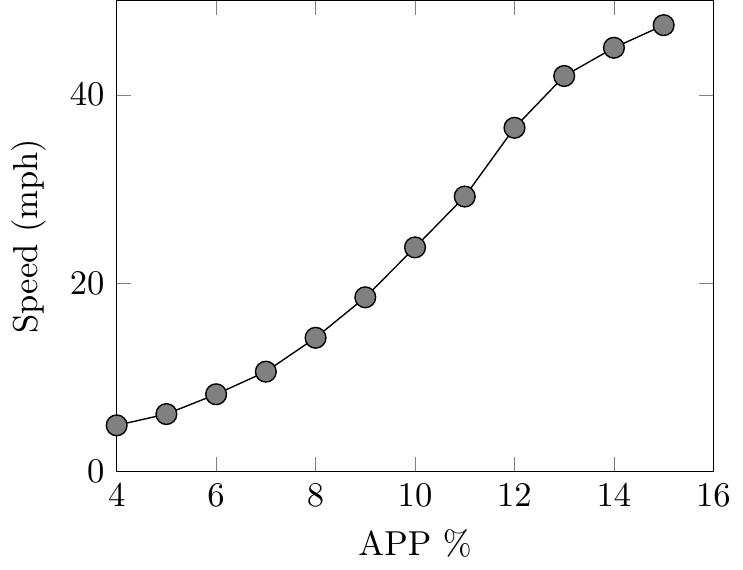}\hfill
	\label{f:maxscatter}
	}
	\subfloat[]{
		\centering
	\includegraphics[width=0.24\textwidth]{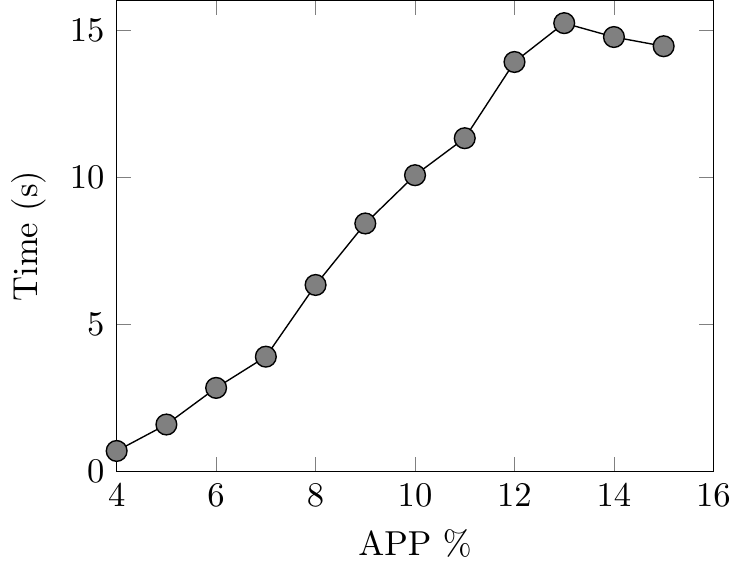}
	\label{f:apptau}
	}
	\subfloat[]{
			\centering
			\includegraphics[width=0.24\textwidth]{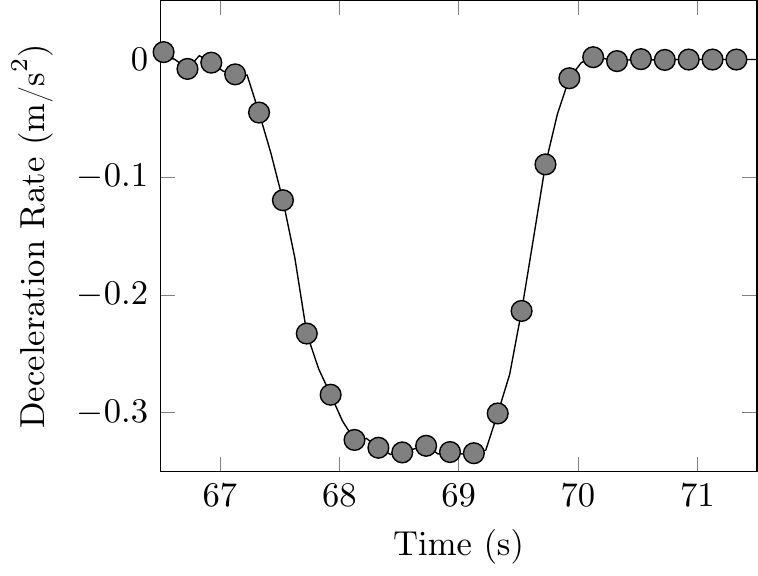}
			\label{f:decel}
		}
	
	\caption{\protect\subref{f:appstep} APP step response for 5\%, 10\%, and 15\% pedal presses. The graph shows a general first order speed response for a given pedal percentage. \protect\subref{f:maxscatter} Vehicle settling speeds for given APP step input percentages. \protect\subref{f:apptau} Time constants for given APP step input percentages. \protect\subref{f:decel} Deceleration rate for BPP step input of 15\%. The figure shows a first order relationship between BPP percentage and deceleration rate.} 
\end{figure*}

For the APP sensor system identification, the vehicle was placed on a dynamometer \cite{mustang} and step inputs were initiated on the APP sensor from 4\% to 15\% at increments of 1\%. Fig. \ref{f:appstep} shows the step responses for some accelerator pedal inputs. It was observed that for a given APP percentage the vehicle would eventually settle at a specific speed. The relationship between APP and speed can be described by a first order transfer function.

\par
 The general equation for a first order transfer function, $G(s)$, can be represented by \begin{equation}
 \label{e:firstorder}
 G(s) = \frac{K}{\tau s + 1}.
 \end{equation} 
 Where \textit{K} is the constant or equation that relates APP to vehicle speed, and $\tau$ is the system time constant. The equation for \textit{K} was derived from a linear fit of the a scatter plot of max speeds from the step input, as shown in Fig. \ref{f:maxscatter}, and given by 
  \begin{equation} \label{e:app}
 f(x) = 3.65x - 9.7.
 \end{equation}
 Where $f(x)$ is the vehicle speed and $x$ is the APP percentage. The test track (shown in Fig. \ref{f:evr}) where the vehicle was operating is an oval track with sharp corners on the north and south side. The sharp corners and the short straightaways limit the vehicle operating speeds to between 15 and 25 mph for the initial automation. The $\tau$ value that best represented the vehicle response between 15 and 25 mph was chosen as the time constant for accelerator pedal input in the longitudinal model. Fig. \ref{f:apptau} shows the time constants for varying APP percentages. The time constant for the accelerator pedal input was chosen to be 7 seconds, as this best represented the system response for the nominal operating conditions.



\begin{figure*}[t]
	\centering
	\subfloat[]{
		\centering
	\includegraphics[width=0.24\textwidth]{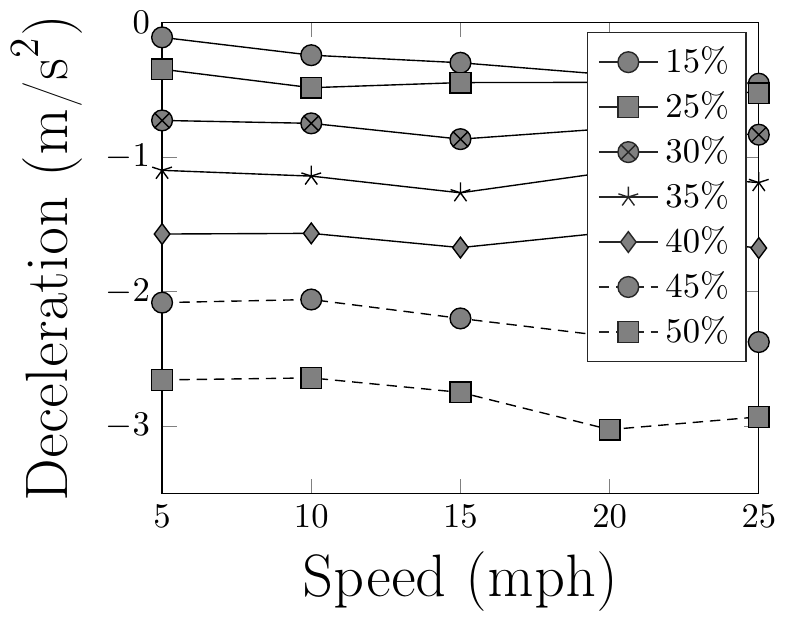}
	\label{f:decelspeed}
	}
	\subfloat[]{
		\centering
	\includegraphics[width=0.24\textwidth]{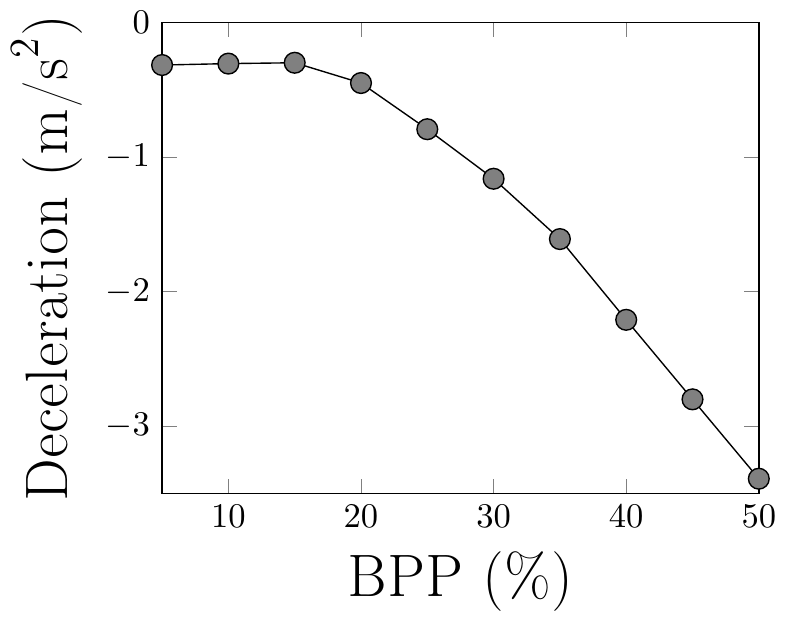}
	\label{f:decelavg}
	}
	\subfloat[]{
			\centering
			\includegraphics[width=0.24\textwidth]{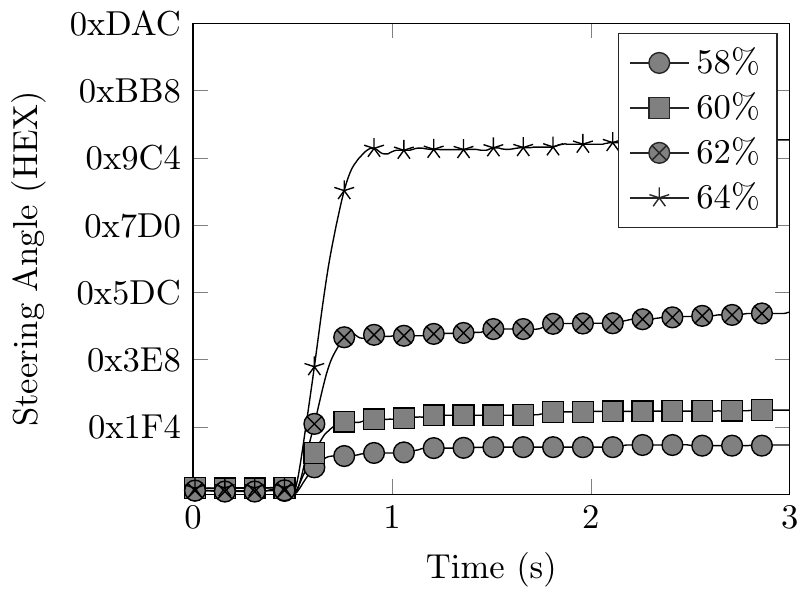}
			\label{f:steerstep25}
		}
	\subfloat[]{
			\centering
			\includegraphics[width=0.24\textwidth]{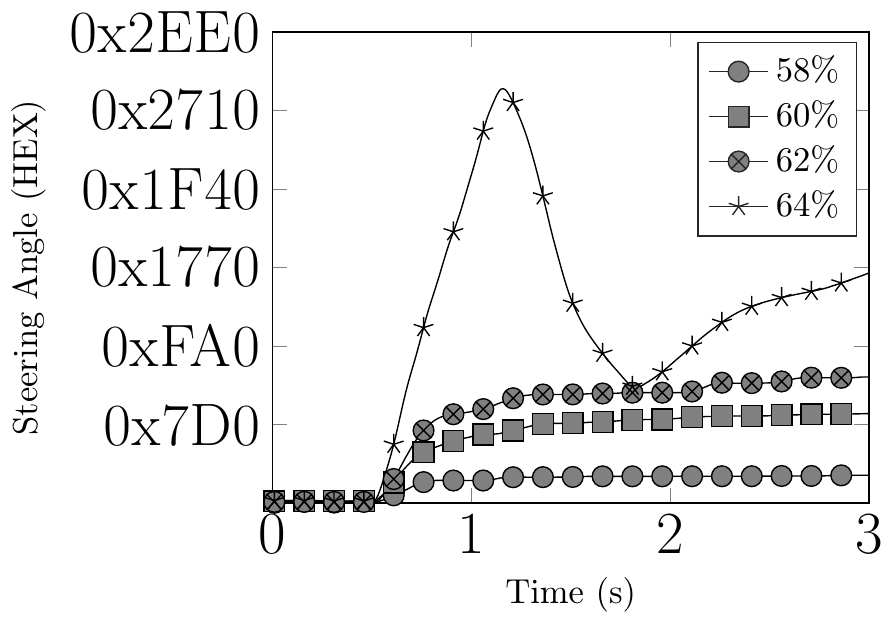}
			\label{f:steerstep15}
		}
	\caption{\protect\subref{f:decelspeed} Vehicle deceleration for BPP percentages at a variety of speeds. The values are the average of the deceleration rates settling point in response to a BPP step input. The lines for BPP percentages do not cross and indicate speed independence for the brake model. \protect\subref{f:decelavg} Average deceleration settling rates due to BPP step input percentages. \protect\subref{f:steerstep25} Steering torque step response for 58\%, 60\%, 62\%, and 64\% duty cycles at a vehicle speed of 25 mph. The plot indicates a first order relationship between torque duty cycle input and steering wheel angle. \protect\subref{f:steerstep15} Steering torque step inputs for 58\%, 60\%, 62\%, and 64\% duty cycles at a vehicle speed of 15 mph. The plot indicates a first order relationship between torque duty cycle input and steering wheel angles, however, at high duty cycle percentages the first order relationship is not valid.  } 
\end{figure*}

For BPP system identification, the vehicle was driven in a large, flat, asphalt area at speeds ranging from 5 to 25 mph at 5 mph increments. The vehicle was accelerated to the desired speed by a driver. Once the vehicle obtained the desired speed, an input to the braking system was initiated through the ROS setup discussed in Section \ref{s:platform}. Step inputs were initiated ranging from 5\% to 50\% of BPP percentage at increments of 5\% for each speed value. The speed data seemed to show a consistent rate of change for a given BPP percentage. To confirm this, the speed data was smoothed using a 5 point moving average, the derivative of the smoothed data was taken by calculating the difference between successive data points, and dividing by the elapsed time between data points. Fig. \ref{f:decel} shows the vehicle deceleration due to a braking event. It was observed that the settling value for the deceleration rate was consistent for a given BPP percentage and varying speeds, which concluded that the longitudinal model was independent of current vehicle speed. This speed independence can be seen in Fig. \ref{f:decelspeed} where each line shows the deceleration rate for a given BPP percentage. At low BPP values the lines converge meaning that deceleration is unaffected by very small brake pedal percentages. However, at higher brake pedal percentages the lines show distinct deceleration rates regardless of the vehicle speed. To show the relationship between BPP percentage and deceleration, an average was taken for each BPP value across each of the speeds. The result of this operation is shown in Fig. \ref{f:decelavg}.

\par
Similar to the APP model, the relationship between BPP and deceleration could be described by a first order transfer function. After analyzing the deceleration curves at different BPP percentages and for different speeds the system time constant, $\tau$, was calculated to be 0.3 seconds. The equation that relates BPP to deceleration was determined by finding a curve fit algorithm for the curve in Fig. \ref{f:decelavg}. This would result in an equation that would provide a BPP percentage for a desired deceleration rate. The equation for \textit{K} is given by

\begin{equation}\label{e:bpp}
f(x) = -0.0018x^2 + 0.029x - 0.3768,
\end{equation}
 where $f(x)$ is the deceleration, and $x$ is the BPP percentage. This equation is used to describe \textit{K} from the general first order transfer function equation.


\subsection{Lateral Model}
The lateral model of the vehicle was determined by step response analysis. The model relates an input from the steering torque sensor to changes in the steering wheel angle. As discussed in Section \ref{ss:sensoremulation}, the torque sensor measures the torque applied by the driver, and sends that information to the PSCM. The PSCM activates the power assist motor that connects to the steering rack, and moves the wheels. The steering wheel angle is measured by a sensor in the steering wheel and output on the CAN bus at a high level of precision.
\par
Step inputs were initiated on the steering torque duty cycle signal ranging from 50\% to 63\% at 1\% increments. Tests were performed at a large, flat, asphalt area with vehicle speeds ranging from 5 mph to 25 mph. Fig. \ref{f:steerstep25} shows the results of the step input tests performed at 25 mph. It was observed that a general first order transfer function could be used to describe the relationship between steering torque duty cycle and steering wheel angle. However, at lower speeds and higher torque values this observation is not valid. Fig. \ref{f:steerstep15} shows the step response of the steering system at 15 mph. At the higher torque values the steering wheel angles do not settle to a consistent steering wheel angle. It was also observed that the settling angles for a given steering torque duty cycle are not consistent for varying speeds. Therefore, the lateral model identification is speed dependent and would require a speed dependent limit on the steering torque duty cycle. Providing these characteristics, the system can still be modeled as first order transfer function for a given speed.

\begin{figure*}[t]
	\centering
	\subfloat[]{
		\centering
	\includegraphics[width=0.33\textwidth]{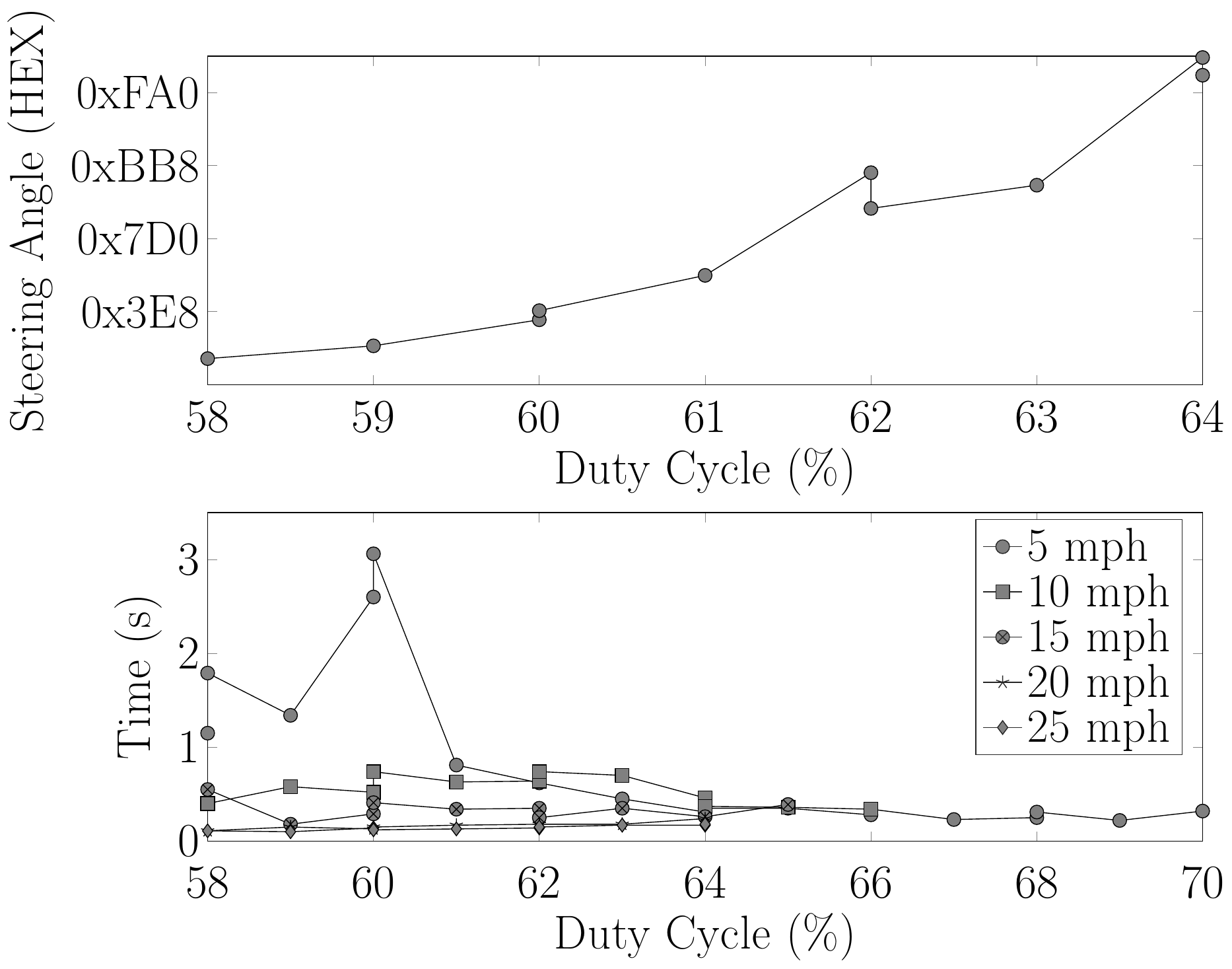}
	\label{f:steermaxtau}
	}
	\subfloat[]{
		\centering
			\includegraphics[width=0.32\textwidth]{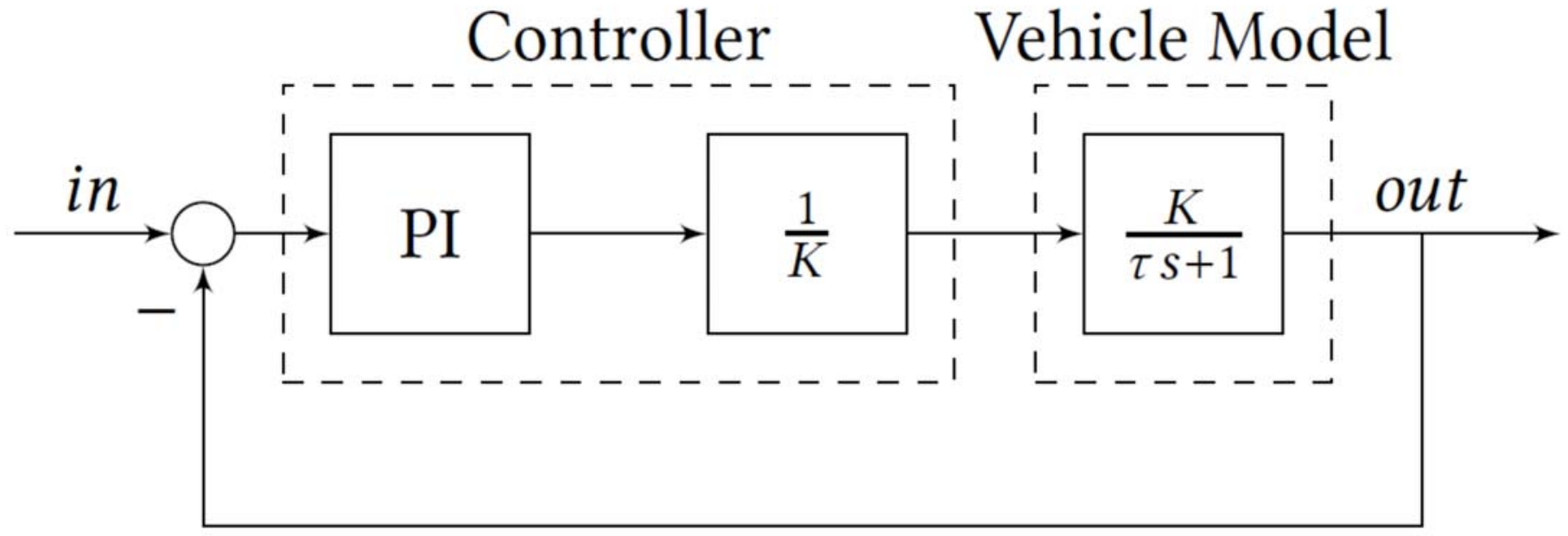}
		\label{f:firstorderloop}
	}
	\subfloat[]{
		\centering
		\includegraphics[width=0.32\textwidth]{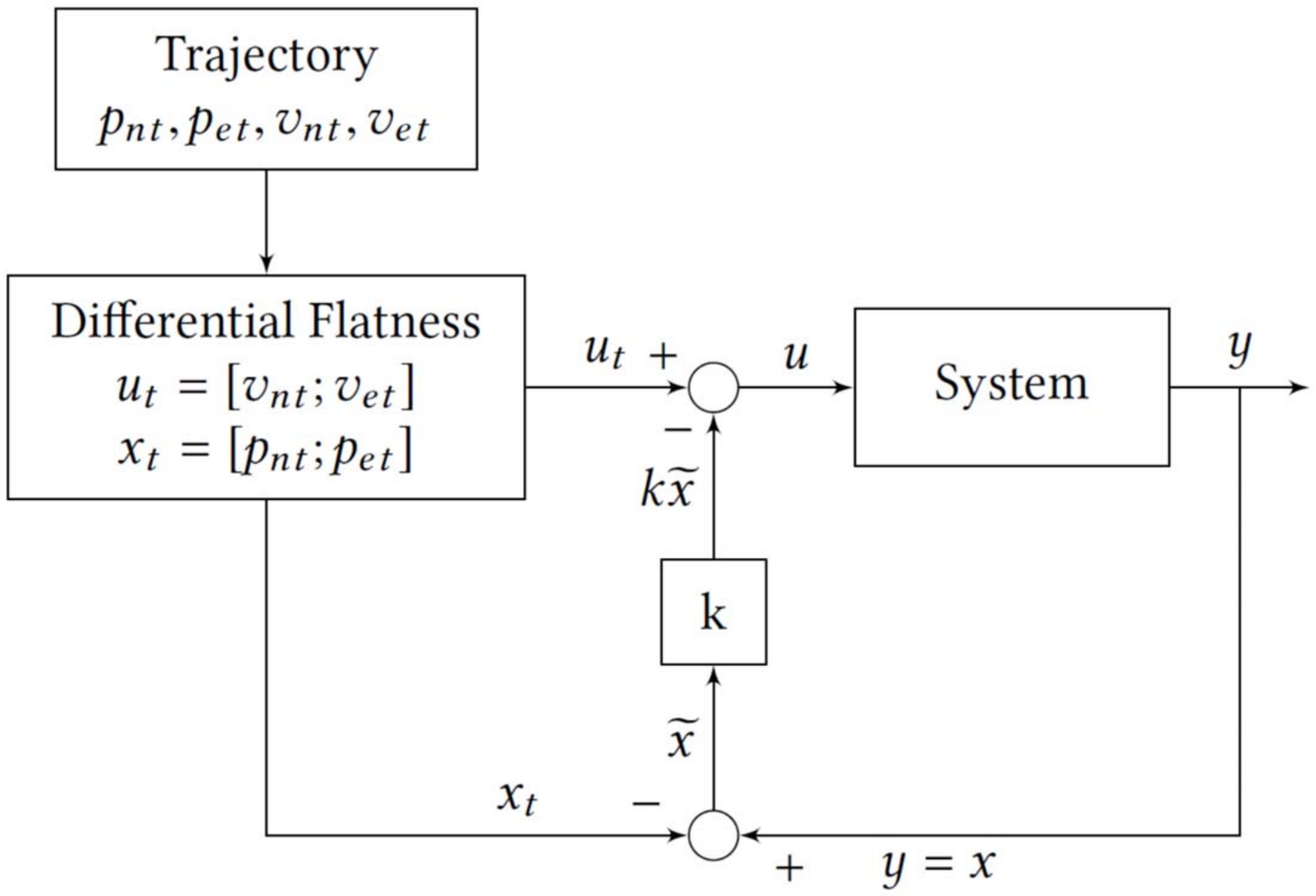}
		\label{f:control_loop}
	}
	\caption{\protect\subref{f:steermaxtau} \textit{Top:} Steering angle settling values for given steering torque duty cycle step inputs. Values represented in hexadecimal format as received from CAN bus. \textit{Bottom:} Steering angle time constants for given steering torque duty cycle step inputs at varying speeds. The time constants converge at higher speeds. \protect\subref{f:firstorderloop} General control loop for a first order PI controller. \protect\subref{f:control_loop} High-level system block diagram. Details the high-level control loop, and the path following algorithm. The system block contains the low-level control loops, and the vehicle.}
\end{figure*}



\par
The steering data was analyzed in order to determine the gain equation, \textit{K}, and the time constant, $\tau$. Time constants were calculated for each step input response and for each speed. Fig. \ref{f:steermaxtau} \textit{Bottom} shows the time constants for given steering torque duty cycles. Each of the lines indicates the speed at which the test was performed. It can be seen that at low speeds and low duty cycles the time constants are not consistent. But at higher speeds the inconsistencies lessen. A time constant, $\tau$, of 0.2 seconds was chosen to optimize for typical vehicle operation.
\par
Since the lateral system was found to be speed dependent, the gain equation \textit{K} must also be speed dependent. The step input tests were performed at 5 mph increments so a gain equation \textit{K} would be found for each speed value. These gain equations relate steering torque duty cycle to steering wheel angle. Fig. \ref{f:steermaxtau} shows the settling angles for varying steering torque duty cycles when the vehicle was traveling at 25 mph. A curve fit approximation was completed for this data set, and a solution was determined by solving the given equation. For this data set, the given equation for K is 
\begin{equation}\label{e:steer}
f(x) = 59.4x^2 - 6802.7x + 195084.5,
\end{equation} 
where $f(x)$ is the steering wheel angle and $x$ is the steering torque duty cycle. 
\par 
Figs. \ref{f:steerstep25} and \ref{f:steerstep15} show the step response of the vehicle due to steering torque input signals. The graphs do not include step input values below 58\% because the step responses at such values had little effect on the steering wheel angle. This exposed a deadband in the response from the steering torque sensor input to the steering wheel angle. A deadband compensation algorithm was implemented to mitigate the effects of this non-linearity. As shown in Fig. \ref{f:high_level_diagram}, the deadband compensation code was executed  just before the signal was sent to the vehicle. If the torque input value was greater than 50\%, then 
\begin{equation}
	\tau_{cmd} = B_{max} + \frac{\tau - 50}{\tau_{max} - 50} \left(\tau_{max} - B_{max}\right)
\end{equation}
was used to compensate for the deadband. If the torque input value was less than 50\%, then
\begin{equation}
	\tau_{cmd} = B_{min} + \frac{50 - \tau}{50 - \tau_{min}} \left(\tau_{min} - B_{min}\right)
\end{equation}
was used to compensate for the deadband. Where $\tau_{cmd}$ is the torque command sent to the vehicle, $\tau$ is the value received from the PI controller, $B_{max}$ is the upper limit of the deadband, $B_{min}$ is the lower limit of the deadband, $\tau_{max}$ is the maximum allowed value for the steering torque signal, and $\tau_{min}$ is the minimum allowed steering torque signal. For the deadband on the 2013 Ford Focus EV, the upper and lower limits were 55\% and 45\%, and the maximum and minimum values for the torque signal were 64\% and 37\%, respectively.


\subsection{PI Controller Design}\label{s:PI}
Low-level control loops were designed to control vehicle speed and steering wheel angle. The desired speed and desired steering wheel angle would be input to the low-level control loops from a user or high-level controller. The low-level longitudinal controller interfaced with the accelerator and brake pedals to effect vehicle speed. A separate loop was designed for each vehicle input, and switching logic was used to choose whether the acceleration or brake loop would be used. The low-level lateral controller would receive the desired steering wheel angle and determine the appropriate input to the steering torque sensor to achieve the desired angle.
\par 
A Proportional Integral (PI) Feedback Controller was implemented for longitudinal and lateral control. Fig. \ref{f:firstorderloop} shows a basic PI Feedback Controller for a first order system. The transfer function block represents the vehicle and contains the system model. The $\frac{1}{K}$ block effectively cancels out the gain equation \textit{K}, and helps relate the speed error to a vehicle input. For example, in the longitudinal controller, the \textit{K} equation receives the APP as an input, and outputs speed. Therefore, the input to the transfer function block must be an APP value. However, the control loop is calculating a speed error, so the output of the PI block is a speed value. The $\frac{1}{K}$ block translates the speed value into appropriate APP value. 

\par
 Since the \textit{K} and $\frac{1}{K}$ can be combined to equal 1, they can be ignored in the loop equation. The open loop transfer function of this system is then given by 
 \begin{equation} \label{e:openloop}
 G_{OL}(s) = \frac{1}{\left(\tau s + 1 \right)}.
 \end{equation}
 Closing the feedback loop and adding the PI controller gives
\begin{equation} \label{e:closedloop}
G_{CL}(s) = \frac{\frac{k_p}{\tau} \left( s + \frac{k_i}{k_p} \right) }{s^2+ \left( \frac{1}{\tau} + \frac{kp}{\tau} \right) s+\frac{k_i}{\tau}}.
\end{equation} 
The system is stable if the real part of the closed-loop poles are negative. Solving for the closed loop poles and zero yields 
\begin{equation}
\label{e:poles}
s = \frac{-\left(kp + 1\right) \pm \sqrt{\left(k_p+1\right)^2 - 4k_i\tau}}{2\tau},
\end{equation}
and
\begin{equation}
\label{e:zero}
s = -\frac{k_i}{k_p},
\end{equation} respectively. From these equations it can be determined that if $k_p$, and $k_i$ are positive the system will be stable.

\par
The second order transfer function obtained through closing the loop can be written, in a general form, as 
\begin{equation}
\label{e:secondorder}
\frac{\omega_n^2}{s^2+2\zeta\omega_n s + \omega_n^2}.
\end{equation} 
Where $\zeta$ is the damping coefficient and $\omega_n$ is the natural frequency. The damping coefficient determines whether the system will be underdamped, overdamped, or critically damped and the natural frequency helps to determine the time constant for the system. The time constant, $\tau$, is given by the equation \begin{math} \tau = \frac{1}{\zeta \omega_n}\end{math}. Values were chosen for the damping coefficient and the time constant to define the system behavior. From these values one can determine the appropriate $k_p$ and $k_i$ for the system. The equations for $k_p$ and $k_i$ are given by
\begin{equation} \label{e:kp}
k_p = \tau\left(2\zeta\omega_n - \frac{1}{\tau}\right),
\end{equation}
and
\begin{equation}\label{e:ki}
k_i = \tau\omega_n^2,
\end{equation}
 respectively. Table \ref{t:loops} shows the calculated values for each of the control inputs, where the $\tau_{car}$ column shows the time constants found during model identification and are internal vehicle parameters.

\begin{table}[b]
	\centering
	
	\begin{tabular}{c|c|c|c|c|c|c}
		\textbf{Input}    & \textbf{$\tau_{car}$} & \textbf{$\zeta$} & \textbf{$\tau$} & \textbf{$\omega_n$} & \textbf{$k_p$} & \textbf{$k_i$} \\ \hline
		Accelerator Pedal & 7                     & 1                & 0.5             & 2                   & 27             & 28             \\
		Brake Pedal       & 0.3                   & 1                & 0.5             & 2                   & 0.2            & 1.2            \\
		Steering Torque   & 0.2                   & 1                & 0.33            & 3                   & 0.2            & 1.8           
	\end{tabular}
	\caption{Table of Values for Input Loops}
	\label{t:loops}
\end{table}

\section{GPS-based High Level Control}\label{s:path}
The high level controller was designed to take a desired trajectory or path, and provide appropriate inputs for the low-level controllers. There are a number of high level control strategies for manned and unmanned vehicle control \cite{medagoda1} \cite{coulter1} \cite{villagra1}. The control strategy should be determined by the system characteristics and the system objectives. This platform was to be used on a ground vehicle to track a desired trajectory and was determined to be differentially flat \cite{Rigatos1} for the chosen states. A differentially flat system is one in which the output is a function of the system states, the input, and the derivatives of the input, and both the states and the input are functions of the output and the derivatives of the output. Therefore, a simple high level controller using the properties of a differentially flat system, and state feedback control were chosen \cite{diff1}. An example control system design for a differently flat Unmanned Aerial Vehicle (UAV) system was demonstrated in \cite{ferrin11}.  The following paragraphs discuss differential flatness, and the high-level controller architecture shown in Fig. \ref{f:control_loop}.

%


\subsection{Differential Flatness}
A system is said to be differentially flat if there exists an output vector $y$ in the form of \begin{equation}\label{e:diff_output} y = h \left( x,u,\dot{u},\ddot{u},....,u^{(r)}\right), \end{equation} such that, \begin{equation} \label{e:diff_states_inputs} \begin{split} & x = \phi\left(y,\dot{y},\ddot{y},....,y^{(r)}\right), \\ & u = \alpha\left(y,\dot{y},\ddot{y},....,y^{(r)}\right), \end{split} \end{equation} where $h$, $\phi$, and $\alpha$ are smooth functions. The output of the differentially flat system can, therefore, be defined by a function of the system states, the control input, $u$, and the derivatives of $u$. The state vector and control input vector can each be described by a function of the flat output, $y$, and its derivatives. 
\par
The state vector for this system was chosen to be 
\begin{equation}
\label{e:x}
x = \begin{bmatrix}
p_n \\
p_e
\end{bmatrix},
\end{equation} 
where $p_n$ is the vehicle position in the north direction and $p_e$ is the vehicle position in the east direction. The system input, $u$, is given by
\begin{equation}
\label{e:u}
u = \begin{bmatrix}
v_n \\
v_e
\end{bmatrix},
\end{equation} 
where $v_n$ is the velocity in the north direction and $v_e$ is the velocity in the east direction. The system output, $y$, is given by
\begin{equation}
\label{e:y}
y = \begin{bmatrix}
p_n \\
p_e
\end{bmatrix}.
\end{equation} 
From equations \ref{e:x}--\ref{e:y} it can be concluded that this is a differentially flat system and can be defined by equations \ref{e:diff_output} and \ref{e:diff_states_inputs}. The dynamic model of the system is then given by 
\begin{equation}
\begin{split}
\label{e:xdot}
\dot{x} &= Ax + Bu
 = \begin{bmatrix}
0 &0\\
0 & 0
\end{bmatrix}
\begin{bmatrix}
p_n \\
p_e
\end{bmatrix}
+
\begin{bmatrix}
1 &0\\
0 & 1
\end{bmatrix}
\begin{bmatrix}
v_n \\
v_e
\end{bmatrix},
\end{split}
\end{equation}
and the output, measured by an RTK-GPS system, is given by
\begin{equation}
\begin{split}
\label{e:yeq}
y &= Cx 
 = \begin{bmatrix}
1 &0\\
0 & 1
\end{bmatrix}
\begin{bmatrix}
p_n \\
p_e
\end{bmatrix}.
\end{split}
\end{equation}

\subsection{Control Architecture}
A virtual target scheme was selected for path and trajectory control. The virtual target data was gathered by driving the vehicle around the test track and recording the RTK-GPS data at 10Hz with 2 cm resolution. The recorded data contained the $p_n$, $p_e$, $v_n$, and $v_e$ data and was replayed in the system as a virtual target. The state, input, and output vectors of the virtual target are denoted as $x_t$, $u_t$, and $y_t$ respectively and $p_{nt}$, $p_{et}$, $v_{nt}$, and $v_{et}$ are the positions and velocities of the virtual target. The goal of the virtual target scheme is to minimize the difference between the system states and the virtual target states. The difference between the vehicle states and target states is given by $\widetilde{x} = x - x_t$, and the input difference is given by $\widetilde{u} = u - u_t$. Therefore, the error model is given by $\dot{\widetilde{x}} = A\widetilde{x} + B\widetilde{u}$, where $u$ is the commanded velocity vector. Solving for $u$, and using a state feedback control strategy, where $\widetilde{u} = -k\widetilde{x}$ gives $u = u_t - k\widetilde{x}$.

The Linear Quadratic Regulator (LQR) method was used to find the optimal gain value for $k$. The entries in the system input vector, $u$, are notated by, $u_1$ and $u_2$, and are used to obtain desired velocity by $v_d = \sqrt{u_1^2 + u_2^2}$, and desired heading by $\psi_d = \tan^{-1}{\left(\frac{u_1}{u_2}\right)}$. Through calculation and experimentation a relationship was determined between the desired heading, $\psi_d$, and the desired steering wheel angle, $\theta_d$. Such that, $\theta_d = k_d\psi_d$. The desired steering wheel angle and the desired velocity are sent to the low level controllers discussed in Section \ref{s:lowlevel}.
\section{Testbed Platform Overview} \label{s:platform}
\begin{figure}[t]
	\centering
	\includegraphics[width=0.49\textwidth]{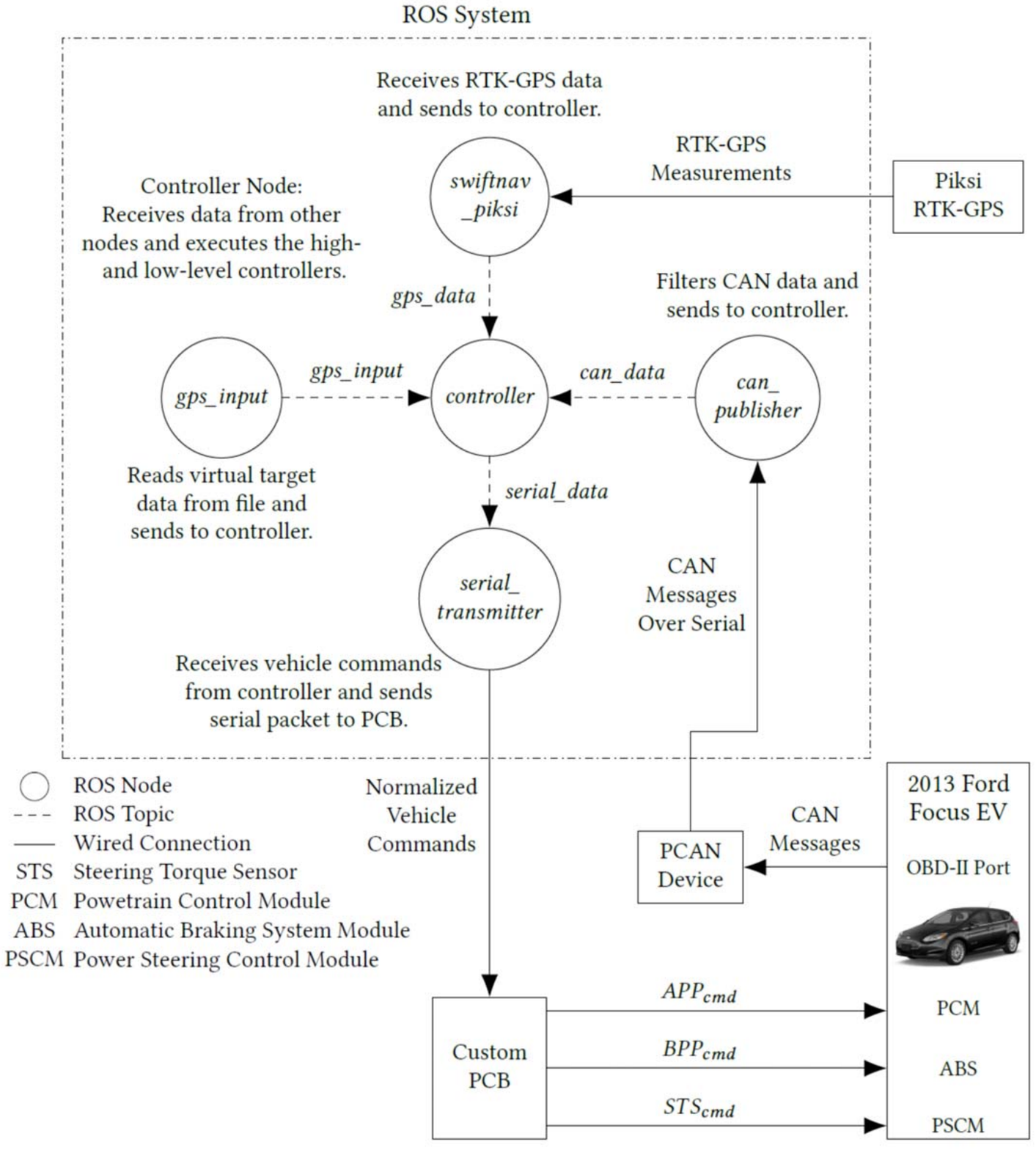}
	\caption{Testbed platform diagram including ROS system, hardware device interfaces, and vehicle interfaces.}
	\label{f:platform_diagram}
\end{figure}

A versatile and robust testbed platform is required to enable full-sized autonomous vehicle research. The platform was designed to enable vehicle automation for both the CAN injection and the sensor emulation approaches discussed in Section \ref{s:reverse}. For the CAN injection approach the platform was able to monitor the CAN bus and inject the desired packets, whereas for the sensor emulation approach the platform required access to the output lines of the sensors to be emulated. In order to proceed to autonomy the platform had the ability to sense the environment, determine vehicle location, communicate with the vehicle, and monitor the CAN bus. A computer running Ubuntu and the Robot Operating System (ROS) was used to communicate between the platform components. The computer was connected to a microcontroller to allow communication with the vehicle. Fig. \ref{f:platform_diagram} shows a diagram of the autonomous system, including the ROS software architecture, hardware connections to devices, and the vehicle interfacing hardware.
\par
A ROS-based platform was chosen for ease of use, modularity, and sensor interfacing packages. ROS is an open source framework that encourages collaboration between researchers. People can contribute to the ROS effort by creating software packages that interface with common sensors and provide tools for development. For example, an open-source software package for ROS was provided by Swift Navigation to interface with the Piksi RTK-GPS units \cite{piksi}, which helped expedite development time for this project. The ROS architecture components used in this project are packages, nodes, and topics. A ROS package is a collection of executable files used to complete a task. Generally packages are used to compartmentalize similar parts of a project. ROS nodes are the executable files in a ROS package that can be written in C/C++ or Python. A ROS topic is a way to transfer data between nodes. Any node can publish data to a topic, and any node can subscribe to that topic to receive the data. In this sense, the ROS topic acts as a bus to transfer data.
\par 

The following subsections detail the Interfacing Architecture, Sensing Architecture, and the Computational Architecture of the automation platform.

\begin{figure*}[t]
	\centering
	\subfloat[]{
		\centering
		\includegraphics[width=0.32\textwidth]{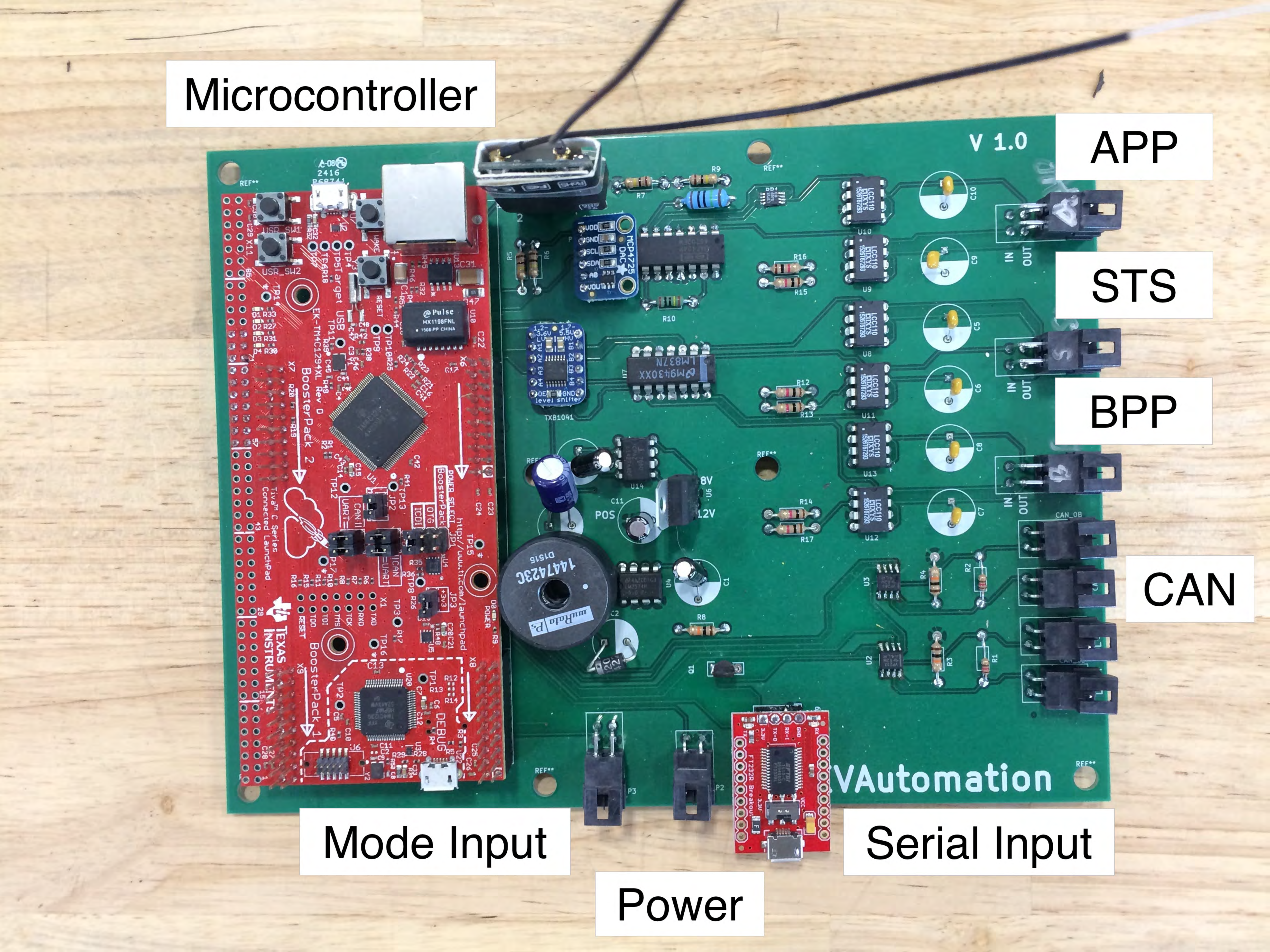}\hfill
		\label{f:pcb}
	}
	\subfloat[]{
		\centering
		\includegraphics[width=0.32\textwidth]{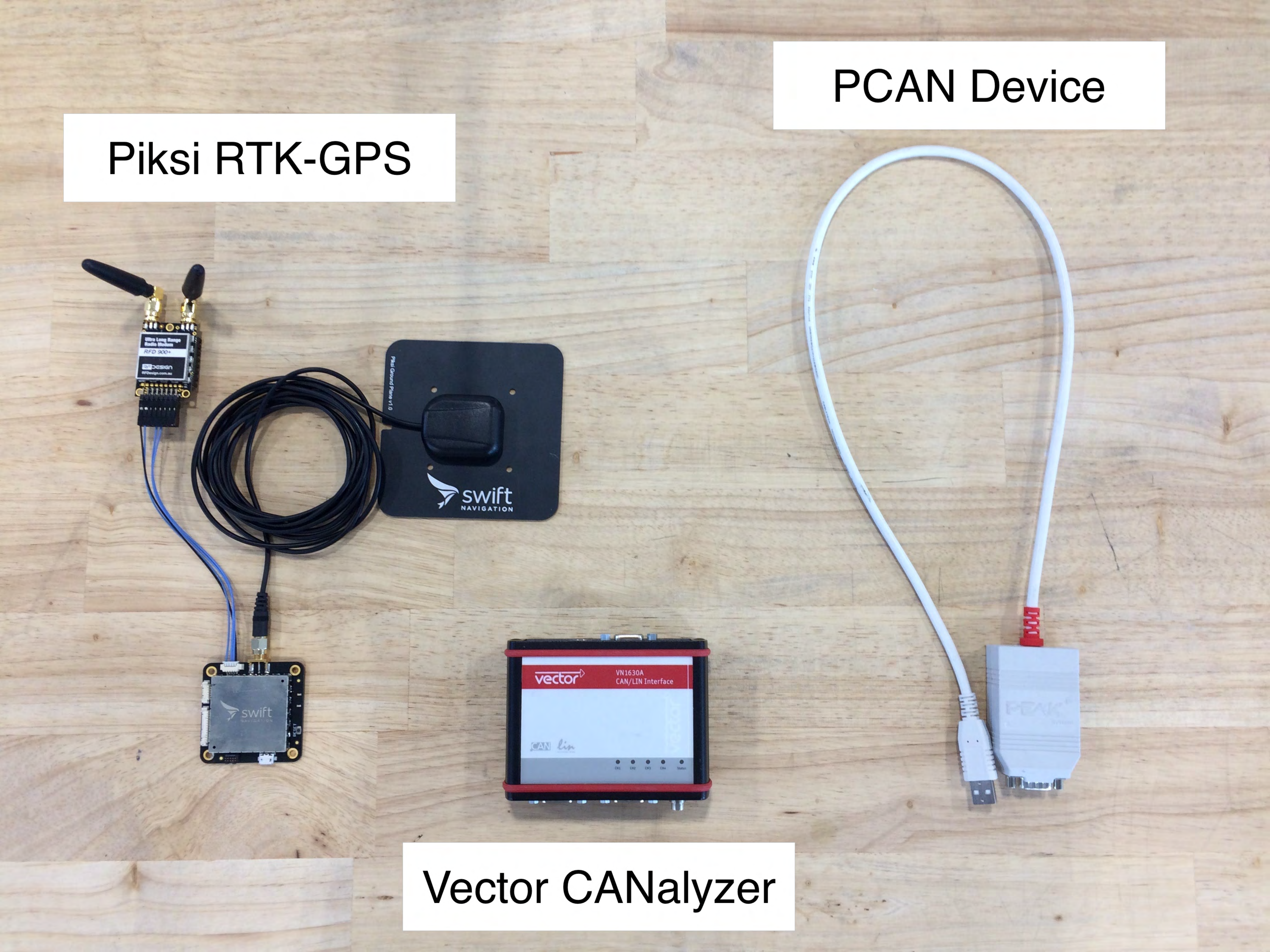}\hfill
		\label{f:piksi_pcan}
	}
	\subfloat[]{
		\centering
		\includegraphics[width=0.32\textwidth]{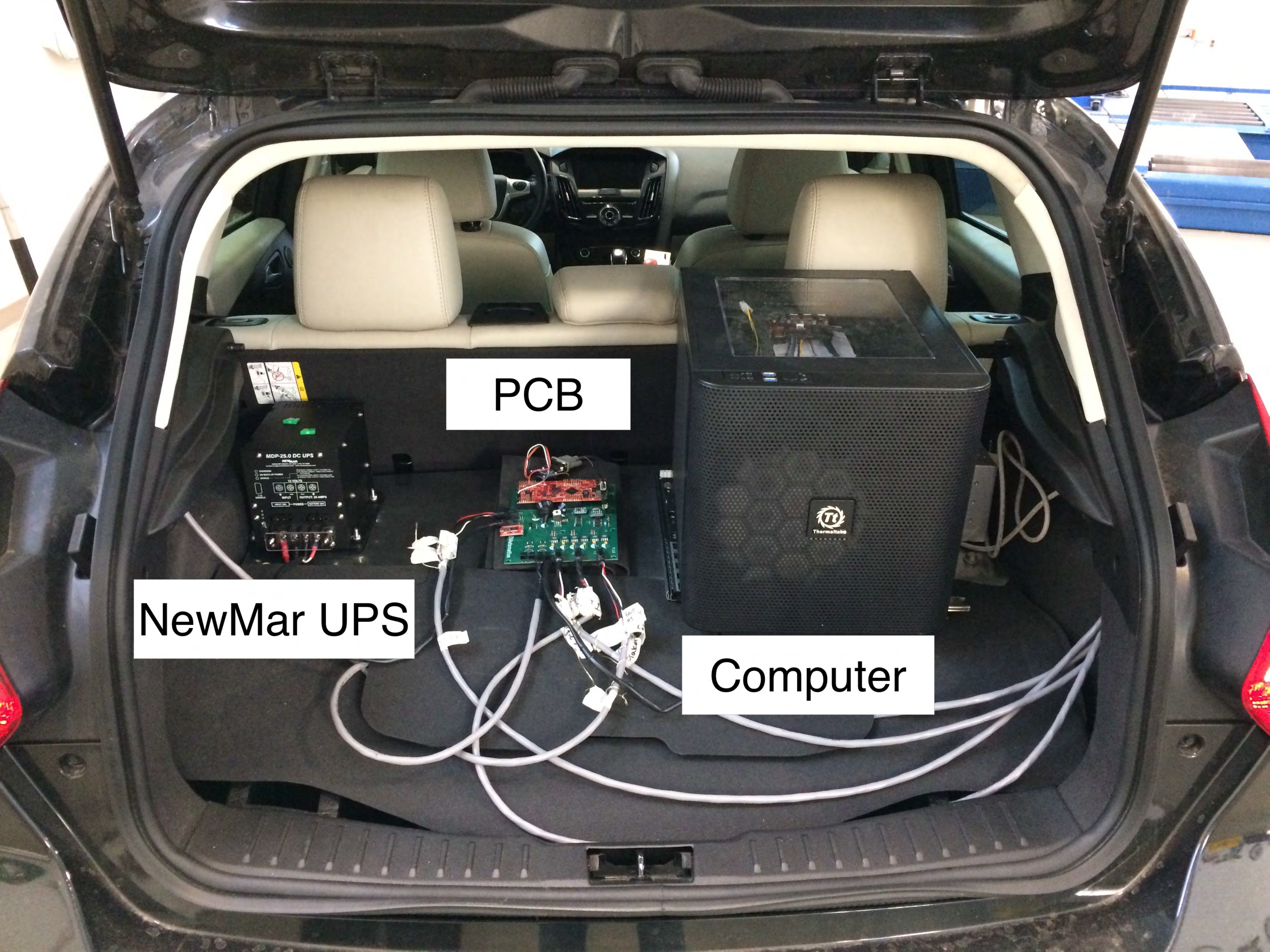}\hfill
		\label{f:trunk}
	}
	
	\caption{\protect\subref{f:pcb} The custom circuit board used to interface with the vehicle CAN bus and generate the signals required for the sensor emulation approach. \protect\subref{f:piksi_pcan} \textit{Left:} The Piksi RTK-GPS unit with GPS antenna, radio antenna for communication, and the Piksi module. \textit{Center:} Vector CANalyzer system, used for reverse engineering CAN messages. \textit{Right:} The PCAN device used to interface with the vehicle CAN bus for high-level control feedback. \protect\subref{f:trunk} The trunk of the 2013 Ford Focus EV with the hardware setup. \textit{Left:} The NewMar UPS that connects to the vehicles 12 V battery and supplies power to the circuit and computer. \textit{Center:} The custom PCB for interfacing with the vehicle CAN bus and sensors. \textit{Right:} Computer running Linux and ROS, connected to the Piksi RTK-GPS unit(s), PCAN Device, and PCB.}

\end{figure*}

\subsection{Interfacing Architecture} \label{ss:interface}
Interfacing devices are critical to the success of vehicle automation as they provide a way to send commands to the vehicle, and monitor the vehicle for feedback. A PCB (shown in Fig. \ref{f:pcb}) was designed to provide a connection between the microcontroller and the vehicle. The following subsections discuss the microcontroller and associated hardware, and the other interfacing devices used for this platform.

\subsubsection{Microcontroller and Associated Hardware}
The TI TM4C129XL evaluation kit was the chosen microcontroller platform because it offered multiple CAN bus interfaces allowing for a combination of CAN injection and sensor emulation from the same board \cite{ti}. The microcontroller receives input from the computer through a UART module. After performing the appropriate computations, PWM signals are generated and appropriate DC voltage levels are determined for vehicle input. The control signals pass through a variety of circuit components to prepare the signals for vehicle injection. The signals are terminated at solid state relays that select either the original vehicle signal, or the generated signal to be sent to the vehicle. The user determines which signal is sent based on a mode switch input to the microcontroller.
\par
The sensor emulation approach requires four PWM signals to be generated by the microcontroller. The signals are passed through a level shifter to shift the amplitudes from 3.3 V to 5 V, and then sent through operational amplifiers in a voltage follower configuration to help drive the signals. The PWM signals are then terminated at the normally opened terminals of solid state relays. 
\par
The accelerator pedal input is generated by a Digital to Analog Converter (DAC) that receives an I2C signal from the microcontroller. The DAC converts the digital communication to an analog voltage level, and sends it to an active filter IC to clean the signal and perform a gain two operation to provide the two output signals. The generated signals then terminate at the normally open terminals of the solid state relays.
\par
Another key feature of the PCB is the safety circuit. Next to the driver there is a mode switch and an Emergency Stop button. The mode switch allows the driver to switch between Manual Driving Mode and Autonomous Mode. The power and solid state relay control signals are routed to the front of the vehicle so the safety driver can switch between operating modes or press the Emergency stop button to prevent power from reaching the circuit. The vehicle's original sensor signals are connected to the normally closed terminals of the solid state relays, so removing the power returns the vehicle to Manual Mode. Power to the circuit is provided by a NewMar DC Uninterruptible Power Supply (UPS) which connects to the 12 V car battery, and provides safe and stable voltage levels for the circuit operations \cite{newmar}. The NewMar UPS also has an internal backup battery. The voltage level is stepped down to $\pm$ 8 V, 5 V and 3.3 V, and distributed across the custom PCB. The UPS is shown in Fig. \ref{f:trunk}. 

\subsubsection{Other Interface Devices}
 The Peak Systems PCAN device was chosen to monitor CAN traffic \cite{pcan}. The PCAN device can connect directly to the vehicle's OBD-II port, and provides serial output over USB. A picture of the PCAN device is shown in Fig. \ref{f:piksi_pcan}. Every message on the connected CAN bus is received and sent serially to the computer. The user can determine which CAN data packets are important to operation, and ignore the rest. One approach to determine necessary CAN data packets is detailed in Section \ref{s:reverse}. Instead of using the CANalyzer system to monitor CAN traffic, the PCAN device can be used to record CAN traffic for a desired event (e.g. vehicle acceleration). The CAN data can be replaced section by section until a message or set of messages is isolated. Additional information on the use of the PCAN device for system feedback is given in Section \ref{ss:ros}.
\par
A USB-to-serial device was used between the microcontroller and computer to enable communication. The control system determines the appropriate inputs to the vehicle and sends the commands to the microcontroller. The device receives the signal from the USB port and sends it to a UART module on the microcontroller. More information about the microcontroller and control system is given in Section \ref{ss:comparch}.
\par 
The TI SN65HVD230 CAN Transceiver Breakout Board \cite{can} \cite{ti2} was used to connect the microcontroller to the CAN bus. This board provided a direct connection with the vehicle CAN bus that can be used for monitoring and injection.

\subsection{Sensing Architecture}\label{ss:sense}
The Swiftnav Piksi RTK-GPS unit \cite{rtk} \cite{kaplan1} was chosen to provide position and velocity estimates for the vehicle. A picture of the Swiftnav Piksi unit is shown in Fig. \ref{f:piksi_pcan}. Swiftnav provides an inexpensive, open-source RTK-GPS solution that provides GPS measurements at 10 Hz and $\pm$2 cm accuracy. In order to achieve such high accuracies the system must have a base station with an RTK-GPS unit, and a second unit mounted to the vehicle. The two units communicate with radio transceivers at 955 MHz. The unit mounted on the car connects to a computer over USB and provides position and velocity data. The base station simply needs a 5 V power supply.
\par 
The 2013 Ford Focus EV has an array of sensors on the vehicle that monitor everything from wheel speed to tire pressure. However, the vehicle does not have high precision wheel encoders, an RTK-GPS receiver, or inertial measurement sensors (IMU's) that could be useful for vehicle automation. The sensor data is typically received by a module and sent on the CAN bus. Using the PCAN device described in Section \ref{ss:interface}, the on-board sensor information can be provided to the rest of the automation platform. For autonomous driving, the accelerator pedal position sensor, brake pedal position sensor, steering wheel angle sensor, and vehicle speed data are used for feedback in the low-level controllers.
\par
The sensor data broadcast on the CAN bus does not provide the information in empirical units, and sometimes the data is masked with other signals. An important aspect to the sensing architecture is the conversion from CAN bus messages to useful units. These conversions could be found experimentally for each message found on the CAN bus, but for the purposes of this testbed the vehicle speed was the only message converted to empirical units (MPH). The vehicle speed is found on the message with arbitration ID \texttt{0x75} on bytes 7 and 8, and is represented by a 16-bit value where byte 7 is the upper byte and byte 8 is the lower byte. When the vehicle was not moving the vehicle speed was represented as \texttt{0xB0D4} on the CAN bus. The decimal representation of this constant, 45,268, is subtracted from the  16-bit vehicle speed to align the 0 MPH value. The vehicle was driven with the RTK-GPS units to provide a reference for the vehicle speed, and it was determined that the CAN value would then need to be divided by 54 to achieve an accurate measure of speed. This process is summarized by 
\begin{equation}
v_{mph} = \frac{(b7 << 8) + b8 - 45268}{54},
\end{equation}
where $b7$ and $b8$ are the integer representations of bytes 7 and 8 from the CAN message with arbitration ID \texttt{0x75}, and the $<<$ operator represents a left bit shift.


\subsection{Computational Architecture}\label{ss:comparch}
The computational architecture includes the code required to combine sensor information, controller commands, and prepare command insertion. The two computational platforms used in this system are the microcontroller and ROS. These platforms are discussed in the following sections and code for these platforms can be found at \url{https://github.com/rajnikant1010/EVAutomation}.

\subsubsection{Microcontroller Software}
The code for the TI TM4C129XL was written in C and took advantage of the built in functionality of the TivaWare Peripheral Driver Library from Texas Instruments \cite{ti3}. Table \ref{t:peripherals} shows the peripherals used and their functionality. The following paragraphs discuss the microcontroller code.
\par 
The microcontroller receives a serial packet from the computer in the form shown in Fig. \ref{f:serial}. The first byte is always \texttt{0xFA}, the second byte gives the number of bytes in the payload, the payload contains the steering, braking and acceleration commands to be sent to the vehicle, and the last two bytes is a 16-bit Cyclic Redundancy Check (CRC) using the CRC16-CCITT algorithm to ensure data integrity. Once received, the CRC is calculated to ensure correct data, and the payload values are stored in appropriate variables. All input commands are normalized between zero and one. For PWM signals the normalized value represents the duty cycle of the signal, where 0.5 represents 50\% duty cycle.

\begin{table}[bp]
	\centering
	
	\begin{tabular}{c|c|c|l}
		\textbf{Port}  & \textbf{Pin} & \textbf{Type} & \multicolumn{1}{c}{\textbf{Purpose}} \\ \hline
		GPIO\_F        & 2            & PWM           & Steering signal 1                    \\
		GPIO\_F        & 3            & PWM           & Steering signal 2                    \\
		GPIO\_F        & 1            & PWM           & Brake signal 1                       \\
		GPIO\_G        & 1            & PWM           & Brake signal 2                       \\
		GPIO\_K (I2C4) & 6            & I2C\_SCL      & Acceleration I2C SCL line            \\
		GPIO\_K (I2C4) & 7            & I2C\_SDA      & Acceleration I2C SDA line            \\
		GPIO\_C        & 4            & Logic         & Mode select signal from user         \\
		GPIO\_C        & 5            & Logic         & Mode signal to system               
	\end{tabular}
	\caption{Microcontroller Peripherals Table}
	\label{t:peripherals}
\end{table}

\begin{figure}
	\centering
	\include{tikzFigures/fig_serial_bytes}
	\caption{Serial message structure for communication with the microcontroller. The serial communication sends commands to the vehicle emulating the APP, BPP, and steering torque sensors. The second byte indicates the number of bytes in the payload, \textit{n}.}
	\label{f:serial}
\end{figure}
\subsubsection{ROS Architecture}\label{ss:ros}
The ROS architecture consists of a series of packages, nodes, and topics \cite{ROS2}. A ROS package can be used to modularize code. For example, the four packages used for this project were \textit{swiftnav\_piksi} (GPS), \textit{focus\_serial} (serial communication), \textit{pcan} (CAN interface),  and \textit{focus\_control} (high- and low-level control for GPS path following). Each of these packages has at least one node and publishes or subscribes to certain topics. A program file (either C/C++ or Python) is written for each node and when a node publishes information to a topic, other nodes can subscribe to that topic and receive the information that was published. The following paragraphs will discuss each of the ROS packages, nodes, and topics used for this system.
\par
The \textit{swiftnav\_piksi} package has only one node. This node initializes a connection with the GPS unit, receives the raw GPS data, and packages the data into an Odometry message. An Odometry message is a standard ROS message that contains information about position, velocity, angular positions, angular velocities, quaternions, and covariance matrices. The Odometry message is published to a topic called \textit{gps\_data}.
\par
The \textit{pcan} package has one node called \textit{can\_publisher} that receives CAN data from the PCAN device and parses the requested data. The CAN data is translated to useful units for the given control strategy, and published to a ROS node called \textit{can\_data}. The \textit{can\_data} topic could provide as much CAN information as the user would like. For the purposes of automating this vehicle, the CAN data of interest is vehicle speed in MPH and steering wheel angle.
\par
The \textit{focus\_control} package has two nodes: \textit{gps\_input} and \textit{controller}. The \textit{gps\_input} node is written in python, and it reads a \texttt{.mat} file with information for a virtual target. The position and velocity data for the virtual target are published to the \textit{gps\_input} topic. The \textit{controller} node subscribes to the \textit{can\_data}, \textit{gps\_data}, and \textit{gps\_input} topics. From the data in these topics it performs a path following algorithm described in Section \ref{s:path}. The path following algorithm provides a desired turn rate (steering wheel angle) and desired velocity for the vehicle.  The low level PI controllers discussed in Section \ref{s:PI} receive the desired values and calculate the appropriate vehicle commands. The accelerator pedal position, brake pedal position, and steering torque duty cycle commands are then published to the \textit{serial\_data} topic. 
\par 
The \textit{focus\_serial} package has one node called \textit{serial\_transmitter}. This node subscribes to the \textit{serial\_data} topic and sends the accelerator pedal position, brake pedal position and steering torque duty cycle information to the microcontroller. Before the data is sent, a Cyclic Redundancy Check (CRC) is performed and a checksum is added to the serial message. The microcontroller checks the CRC to verify the accuracy of the data before sending the requested commands to the vehicle.

\section{Experimental Results} \label{s:results}
Experiments were conducted to determine the results of the autonomous vehicle platform. A video of the system in operation and the development process can be found at \url{https://youtu.be/7ohWIwb6KfM}. The low level controllers were tested with given desired steering angles and velocities. The low level steering controller was improved by implementing a deadband compensation algorithm. The results for the low-level controllers are given in Section \ref{ss:resultsLow}.
\par 
After the low level controllers were verified through testing, experiments were conducted for full vehicle automation by including the high-level path following controller. The automation results are shown in Section \ref{ss:resultsHigh}. 

\subsection{Low Level Controller}\label{ss:resultsLow}
\begin{figure*}[t]
	\centering
	\subfloat[]{
		\centering
			\includegraphics[width=0.33\textwidth]{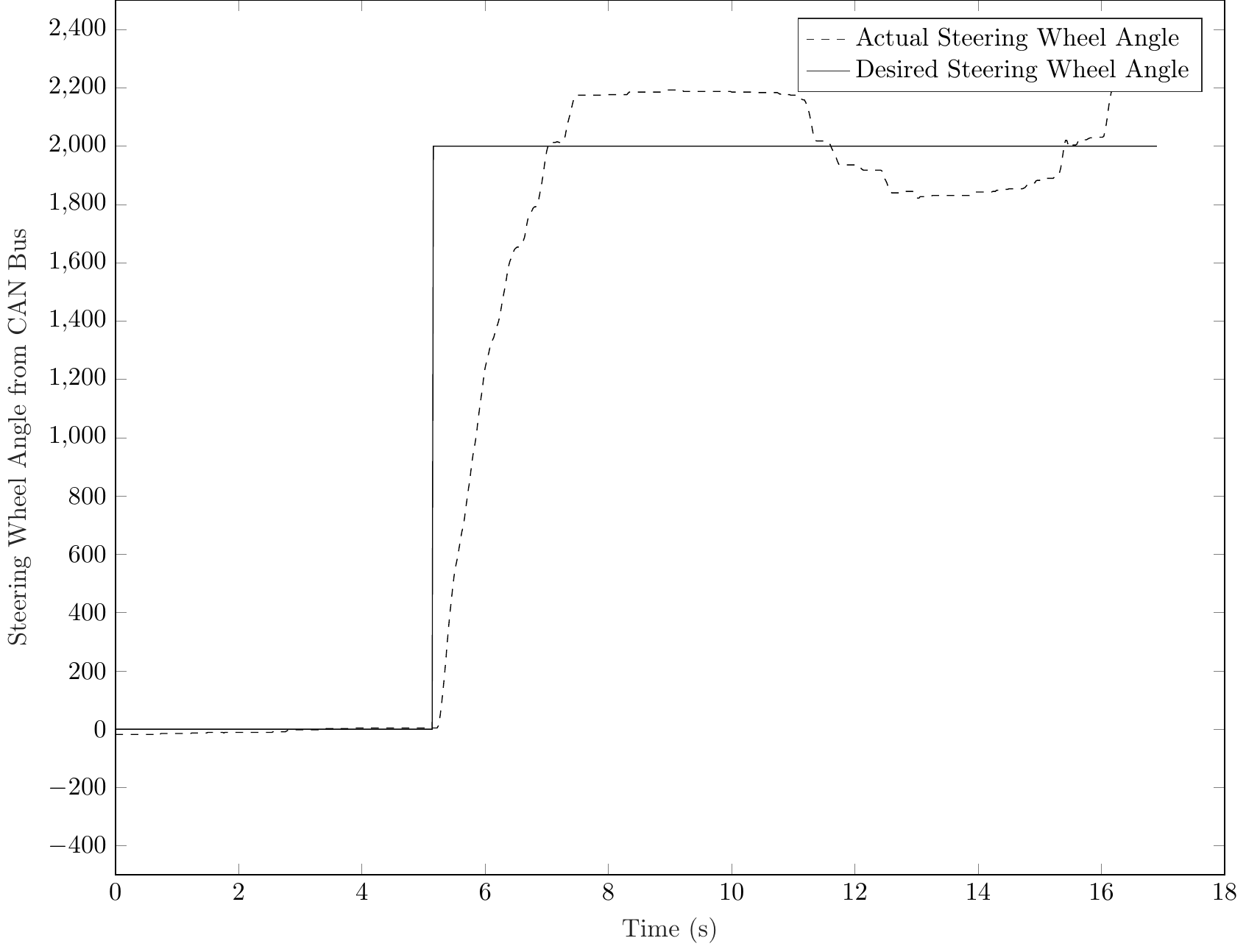}
		\label{f:steerstep}
	}
	\subfloat[]{
		\centering
		\includegraphics[width=0.33\textwidth]{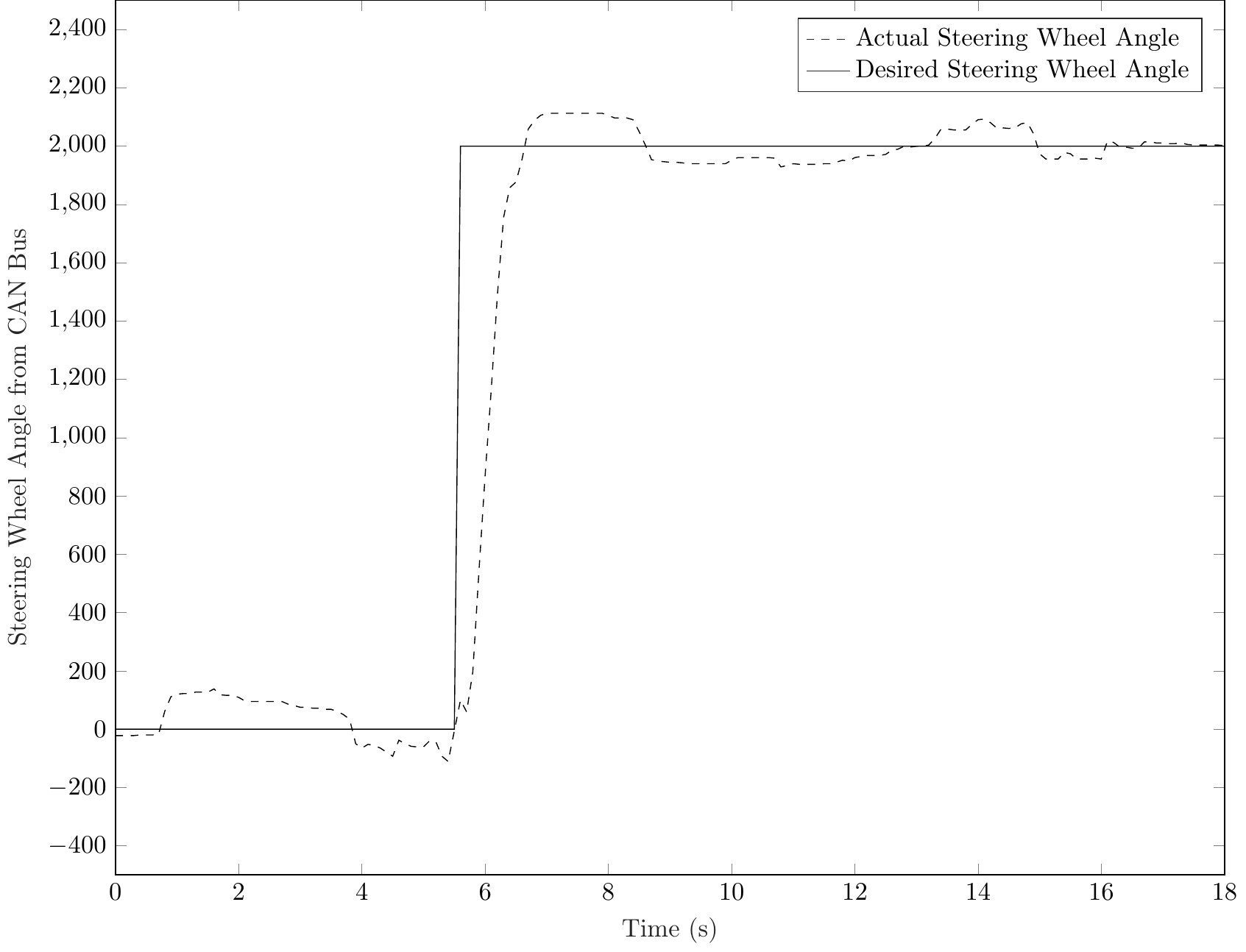}
		\label{f:steerstep2}
	}
		\subfloat[]{
			\centering
			\includegraphics[width=0.33\textwidth]{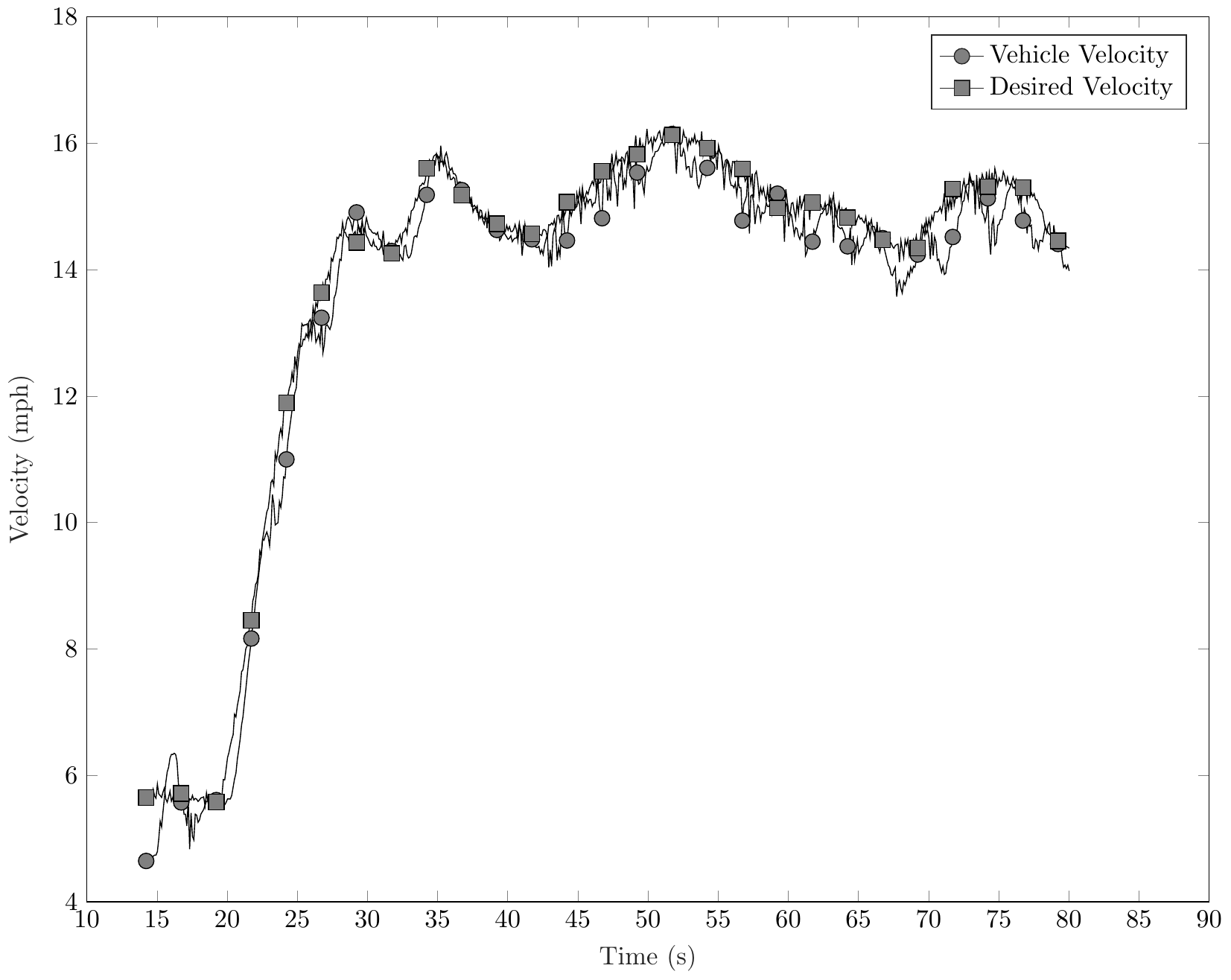}
			\label{f:velocity2}
		}
	
	\caption{\protect\subref{f:steerstep} Steering angle step response without deadband compensation. \protect\subref{f:steerstep2} Steering angle step input with deadband compensation.  \protect\subref{f:velocity2} Desired and actual velocity of autonomous vehicle.}
\end{figure*}

The low level controllers provide speed and steering wheel angle control through the user input signal. For the steering controller, the steering wheel torque sensor signal was used to change the position of the steering wheel. As discussed in Section \ref{s:lowlevel}, the control loop was designed such that, given a desired angle, the controller would change the steering torque value until the desired angle was achieved. This controller was tested using a step input and a graph of the result can be seen in Fig. \ref{f:steerstep}. The y-axis is the steering wheel angle as represented by a Hex value on the CAN bus. A desired steering wheel angle of \texttt{0x7D0} was used as an input to the controller node of the ROS platform. The step response had a maximum value of \texttt{0x891}, representing an overshoot of 9.65\%. The resulting time constant of the system was 1.86 seconds. The desired behavior was for the system to be critically damped and have a time constant of 0.33 seconds.

\par
After implementing deadband compensation, the lateral controller improved. Fig. \ref{f:steerstep2} shows the step response of the lateral controller with deadband compensation. The maximum value for this response is 2,113, which represents a 5.65\% overshoot, and has a time constant of 1.1 seconds.
\par

The result of the longitudinal controller during autonomous driving is shown in Fig. \ref{f:velocity2}. The velocity error is shown in Fig. \ref{f:error_vel_path}, where the average error for the autonomous test was 0.34 mph (0.151 m/s).

\subsection{Path and Velocity Errors of Autonomous Driving}\label{ss:resultsHigh}
The high level controller was tested on the 2013 Ford Focus EV. A recorded data set of the vehicle being driven around the track was used as a virtual target for the differential flatness algorithms discussed in Section \ref{s:path}. The path error was calculated using the euclidean distance from the vehicle to the closest point on the desired path and is given by \begin{equation} d_{er} = \sqrt{\left(p_n - p_{nt})^2 + (p_e - p_{et}\right)^2}. \end{equation}



\par

%

\begin{figure}[t]
	\centering
	\subfloat[]{
		\centering
		\includegraphics[width=0.45\textwidth]{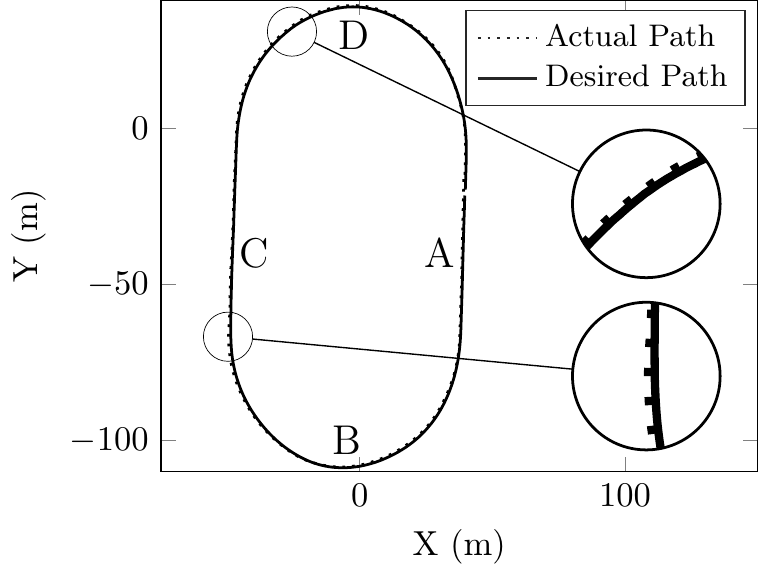}
		\label{f:traj2}
	}
	\subfloat[]{
		\centering
		\includegraphics[width=0.45\textwidth]{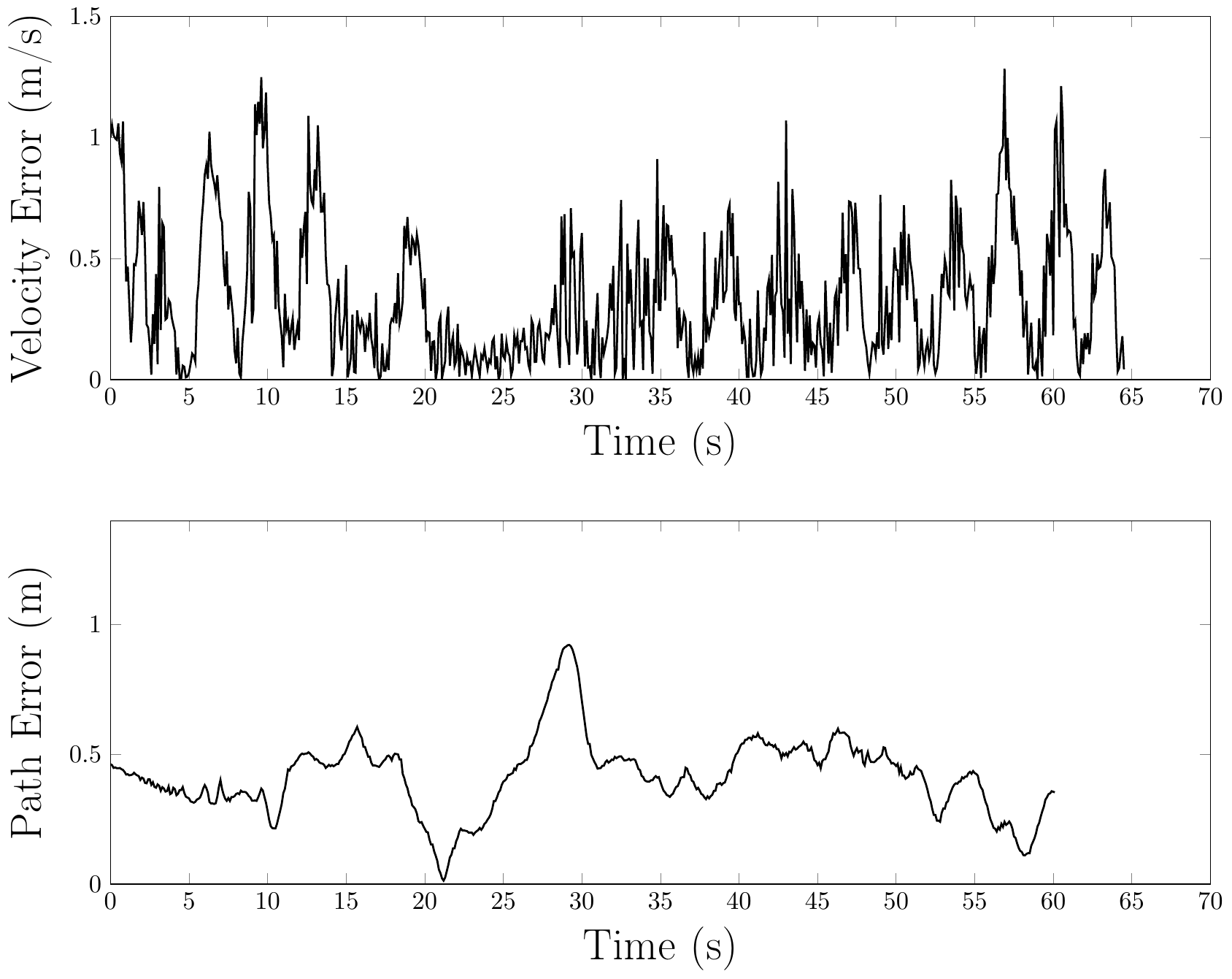}
		\label{f:error_vel_path}
	}
	
	\caption{\protect\subref{f:traj2} Trajectory and path of autonomous vehicle.\protect\subref{f:error_vel_path} \textit{Top:} Velocity error of autonomous vehicle for system. \textit{Bottom:} Path error of autonomous vehicle, as calculated by euclidean distance.}
\end{figure}
A single lap autonomous test was performed with the high- and low-level controllers, and the data was recorded. The trajectory is shown in Fig. \ref{f:traj2}, and the velocity plot is shown in Fig. \ref{f:velocity2}. The path and velocity error plots are shown in Fig. \ref{f:error_vel_path}, where the average path error was 0.43 meters. However, if the vehicle was permitted to immediately perform a second lap of automated driving, the average  path error for the second lap was 0.28 meters, boasting a 34.9\% improvement over the first lap.
\par 


\section{Conclusion}\label{s:conclusion}
The testbed discussed in this work provides an open source vehicle automation system that takes advantage of the vehicle's native communication and control systems to achieve vehicle control. This method was preferred to more expensive and more intrusive platforms on the market today. The overall cost of the components needed to automate a stock vehicle was \$2,315, the itemize cost breakdown can be seen in Table \ref{tab:cost}. The automated systems could be used to further research in a variety of fields, including control systems, vehicle security, and wireless power transfer. 
\begin{table}[b]
	\centering
	
	\begin{tabular}{l|l}
		\textbf{Component}      & \textbf{Cost}    \\ \hline
		Swifnav Piksi RTK-GPS   & \$995            \\
		PCAN - CAN Bus Analyzer & \$225            \\
		TI Tiva C TM4C1294      & \$20             \\
		NewMar UPS              & \$495            \\
		PCB and Components      & \$80             \\
		Computer                  & \$500            \\ \hline
		\textbf{TOTAL}          & \textbf{\$2,315}
	\end{tabular}
	\caption{Itemized Cost Breakdown of Automation Platform}
	\label{tab:cost}
\end{table}
In addition to the GPS-based path following, the low cost automation testbed is to validate a vision based path following algorithm as detailed in~\cite{chase1}. The experiment video of vision based-path following is available at~\url{https://youtu.be/NpAUcNh4QUY}.
\par 
Future work for this project include refining low-level controllers and testing new high-level controllers to improve lateral path accuracy to $\pm$7 cm. To help improve the consistency and avoid errors introduced from GPS, a stereo-vision based controller will be implemented for lane detection.
\section*{Acknowledgment}
The authors would like to thank the EV Automation Team from Utah State University: Cameron Sego, David Petrizze, Hunter Buxton, Tyler Travis, Aaron Kunz, Austin Goddard, Gregory Vernon, Daniel McGary, Clint Ferrin, Ishmaal Erekson, Nicolas Jugganaikloo, Ryker Shirner, and Zachary Garrard. This project would not have been possible without their hard work and dedication to the team goals. The authors would also like to thank the individuals who help manage and coordinate the research at the EVR: Ryan Bohm, Joshua Rambo, and Paul Rau.

\bibliographystyle{IEEEtran}
\bibliography{IEEEabrv,bibliography}

\end{document}